\documentclass[11pt,paper]{article}
\pdfoutput=1
\usepackage{jheppub}

\usepackage[usenames,dvipsnames,table]{xcolor}
\usepackage{graphicx,multirow,array,bm,mathrsfs}
\usepackage{amsmath,amssymb,amsthm}
\usepackage[numbers,sort&compress]{natbib}
\usepackage{cleveref}
\usepackage{bbold}

\usepackage{color}
\usepackage{hyperref}
\hypersetup{hidelinks}
\usepackage{leftidx}
\usepackage{subcaption}

\newtheorem{theorem}{Theorem}

\newcommand{\HH}{{\mathbb{H}}}
\newcommand{\CC}{{\mathbb{C}}}
\newcommand{\RR}{{\mathbb{R}}}
\newcommand{\ZZ}{{\mathbb{Z}}}
\DeclareMathOperator{\hdim}{H.dim}
\DeclareMathOperator{\Tr}{Tr}
\DeclareMathOperator*{\res}{Res}

\renewcommand{\Re}{\operatorname{Re}}

\newcommand{\primes}{\mathcal{P}} 

\newcommand{\op}{\mathcal{O}}
\newcommand{\bdry}{\mathcal{B}}
\newcommand{\bulk}{\mathcal{M}}
\newcommand{\quotient}[1]{#1/\Gamma}
\newcommand{\slice}{\Sigma}

\newcommand\M{\mathcal{M}}

\title{Phase transitions in 3D gravity and fractal dimension}

\author[a,\star]{Xi Dong,}
\note[$\star$]{xidong@ucsb.edu}
\author[b,\wedge]{Shaun Maguire,}
\note[$\wedge$]{smaguire@caltech.edu}
\author[c,\int]{Alexander Maloney,}
\note[$\int$]{maloney@physics.mcgill.ca}
\author[c,d]{Henry Maxfield}
\note[$d$]{maxfieldh@physics.mcgill.ca}

\affiliation[a]{Department of Physics, University of California, Santa Barbara, CA 93106, USA}
\affiliation[b]{Institute for Quantum Information \& Matter, California Institute for Technology, Pasadena, CA, USA}
\affiliation[c]{Department of Physics, McGill University, Montr\'{e}al, QC, Canada.}

\abstract{We show that for three dimensional gravity with higher genus boundary conditions, if the theory possesses a sufficiently light scalar, there is a second order phase transition where the scalar field condenses. This three dimensional version of the holographic superconducting phase transition occurs even though the pure gravity solutions are locally AdS$_3$. This is in addition to the first order Hawking-Page-like phase transitions between different locally AdS$_3$ handlebodies. This implies that the R\'enyi entropies of holographic CFTs will undergo phase transitions as the R\'enyi parameter is varied, as long as the theory possesses a scalar operator which is lighter than a certain critical dimension. We show that this critical dimension has an elegant mathematical interpretation as the Hausdorff dimension of the limit set of a quotient group of AdS$_3$, and use this to compute it, analytically near the boundary of moduli space and numerically in the interior of moduli space. We compare this to a CFT computation generalizing recent work of Belin, Keller and Zadeh, bounding the critical dimension using higher genus conformal blocks, and find a surprisingly good match.}

\usepackage{indentfirst} 

\begin{document}

\maketitle


%


\section{Introduction}

Three dimensional gravity has proven a remarkably rich testing ground for our ideas about classical and quantum gravity.  Even though Einstein gravity possesses no local degrees of freedom, three dimensional theories of gravity nevertheless have many of the rich features of their higher dimensional cousins, including holography \cite{Maldacena:1997re} and black hole solutions \cite{Banados:1992wn} whose Bekenstein-Hawking entropy can be computed microscopically \cite{Strominger:1997eq}.  Theories of gravity in AdS$_3$ are dual to two dimensional conformal field theories, 
allowing one to use CFT methods to gain insight into classical and quantum gravity in AdS.  In this paper we will use CFT methods to motivate the existence of a new class of phase transitions in three dimensional gravity.  We will then verify their existence directly in classical AdS gravity, and explore their features. 

Our central result is simple, and is easiest to state for AdS$_3$ gravity in Euclidean signature.  Such theories can be defined with a variety of boundary conditions: one can take the boundary of (Euclidean) space-time to be any smooth, two dimensional Riemann surface $\bdry$.  
With appropriate boundary conditions \cite{Brown:1986nw} the theory will depend only on the conformal structure of $\bdry$, so can be studied as a function of the conformal structure moduli of $\bdry$.  
The bulk gravity path integral with these boundary conditions is, via AdS/CFT, equal to partition function of the dual CFT on the surface $\bdry$.
We will be interested in gravity theories in the semi-classical limit, where this bulk path integral is dominated by the classical geometry which minimizes the (appropriately regularized) gravitational action.
For example, when the boundary is a sphere the dominant contribution comes from Euclidean AdS$_3$, i.e. hyperbolic space $\HH_3$, which is the unique constant negative curvature metric on the solid ball which ``fills in"  the boundary sphere.   At higher genus, however, 
this path integral can have many saddle points, each of which correspond to a gravitational solution whose boundary is the surface $\bdry$.  For example, when the boundary is a torus the bulk saddles are constant negative curvature metrics on a solid donut which fills in the boundary torus.  There are many such saddles, which are distinguished by which cycle in the boundary torus is contractible in the bulk (see e.g. \cite{Maldacena:1998bw, Dijkgraaf:2000fq, Maloney:2007ud}).  For example, the geometry for which the Euclidean time coordinate is contractible is the (Euclidean) BTZ black hole, while the geometry where the angular coordinate is contractible is interpreted as the ``thermal AdS" geometry used to compute finite temperature observables in a fixed AdS background.    As one varies the moduli of the torus, these two saddles will interchange dominance in the bulk gravity path integral -- this is the three dimensional version of the Hawking-Page phase transition \cite{Hawking:1982dh} describing black hole formation in AdS.

We are interested in the case where the boundary $\bdry$ has genus $g \ge 2$.  Just as in the torus case, there are many different bulk solutions which give saddle point contributions to the partition function, and which can be characterized by a choice of cycles of the boundary surface $\bdry$ which become contractible in the bulk.  The simplest of these solutions are handlebodies, where the bulk solution is the constant negative curvature metric on a solid genus $g$ surface which fills in the boundary $\bdry$.  There will be phase transitions where these geometries interchange dominance: these are the higher genus versions of the Hawking-Page phase transition.  As in the torus case, these handlebodies can be regarded as the Euclidean continuation of AdS$_3$ black holes; they are analytic continuations not of the BTZ black hole, but instead of multi-boundary black holes in AdS
\cite{Brill:1995jv, Aminneborg:1997pz}, as described in \cite{Krasnov:2000zq}.  In the holographic context, these handlebodies describe contributions to the higher genus partition functions of holographic CFTs, which can be used to compute entanglement R\'enyi entropies \cite{Headrick:2010zt}, to constrain OPE coefficients \cite{Cardy:2017qhl, Cho:2017fzo, Keller:2017iql}, or as models of multi-party holographic entanglement \cite{Balasubramanian:2014hda}.
\footnote{There are other ``non-handlebody" solutions as well \cite{Yin:2007gv, Yin:2007at}, which will not concern us in this paper.  Near boundaries of moduli space -- i.e. where cycles in the surface $\bdry$ become small -- the handlebody solutions will always dominate \cite{Yin:2007gv, Yin:2007at}.   Moreover, one can compute numerically the action of the non-handlebodies in the interior of moduli space, and -- at least in the cases which have been studied -- they are always subdominant compared to handlebodies \cite{Maxfield:2016mwh}.
We will therefore focus only on handlebodies in this paper.
}

The bulk solutions described above are all locally $\HH_3$, so can be written as quotients of hyperbolic space of the form $\HH_3/\Gamma$, where $\Gamma$ is a discrete subgroup of the isometry group of $\HH_3$.  Indeed, Einstein gravity in three dimensions has no local degrees of freedom, so any solution of pure Einstein gravity must be locally $\HH_3$.  We are interested in more complicated theories of gravity, however, which have additional degrees of freedom.  In this paper we will consider theories where we have an additional scalar field $\phi$ of mass $m^2$.  This means that the dual CFT has an operator $\cal O$ of dimension $\Delta$, with $m^2 = \Delta (\Delta -2)$.\footnote{We are working in units where the AdS radius is $\ell=1$.}  All of the solutions described above have $\phi=0$, and are dual to CFT configurations on the Riemann surface with $\langle {\cal O} \rangle_\bdry =0$.   

Our central result is the following: in some regions of moduli space, and for sufficiently light scalar fields, the handlebody solutions described above are unstable.  This is because the kinetic operator $(\nabla^2 - m^2)$ will have a negative eigenvalue.  Thus the solution with least action will not be a quotient of AdS, but rather a non-Einstein solutions with $\phi\ne0$.  In the dual CFT, the expectation value of the scalar operator $\langle {\cal O}\rangle_\bdry\ne0$ will be non-zero.  This means that as the moduli are varied there will be phase transitions where these scalar fields condense.
Although the general structure of these $\phi\ne0$ solutions is quite complicated -- we expect the construction of these solutions to be a difficult numerical problem -- we are able to prove rigorously the existence of the instability.

We will see that these phase transitions have several important features, including:
\begin{description}
\item[Instabilities  occur only when the dual operator is sufficiently light.]  
In order for a given handlebody to be unstable, $\Delta$ must be lighter than a certain critical value $\Delta_c$ which we will compute. The value of $\Delta_c$ will depend on the conformal structure moduli as well as on the choice of handlebody.
Whenever the dimension $\Delta<2$ (i.e. the bulk scalar $\phi$ has $m^2<0$) there is some region of moduli space where a given handlebody will be unstable.\footnote{If we require the handlebody to be invariant under a $\ZZ_2$ time reflection symmetry, so that it can be Wick rotated to a real Lorenztian solution, then this condition becomes $\Delta <1$; this would require that the scalar field have $-1< m^2 < 0$ and be quantized with alternate boundary conditions.} 
\item[Instabilities occur only in the interior of moduli space.]  
At the boundary of moduli space one of the handlebody phases will always dominate, and will be stable against the condensation of any scalar field.  In other words, $\Delta_c\to 0$ for the dominant handlebody as we approach the edge of moduli space. As one moves into the interior of moduli space $\Delta_c$ increases so the handlebody becomes more unstable to condensation of the scalar, until the first-order Hawking-Page transition is reached, and a topologically distinct handlebody becomes dominant.

For example, if $\bdry$ is a genus $g=n-1$ surface constructed as an $n$-fold cover of the sphere branched over 4 points, parameterized by their cross-ratio $x$, the handlebody which dominates when $x\to 0$ will become more unstable as $x$ is increased. For this geometry, $\Delta_c$ is a monotonically increasing function of $x$.\footnote{It is important here that we are referring to the handlebody which dominates at small $x$.  For the handlebody which dominates as $x\to 1$, $\Delta_c$ will be monotonically {\it decreasing} function of $x$ between  $0$ and $1$.} 
\item[Handlebodies become more unstable as the genus increases.]  Instabilities only occur when $\bdry$ has genus $g \ge 2$.  If $\bdry$ is the genus $g=n-1$ surface constructed as an $n$-fold cover of the sphere branched over 4 points, then if we hold the cross-ratio $x$ of the four points fixed, the corresponding handlebodies will become more unstable as $n$ is increased.  
In other words, $\Delta_c$ is a monotonically increasing function of $n$. As we take $n\to\infty$ with fixed $x$, $\Delta_c$ approaches a finite value which depends on $x$ but is always greater than $\frac{1}{2}$.
\end{description}

This new phase transition in three dimensional gravity is quite similar to the holographic superconducting phase transition in higher dimensions \cite{Gubser:2008px,Hartnoll:2008vx}.
There is, however, one crucial difference, which is that in the present case the solutions which become unstable are locally AdS$_3$, and have no external potentials (aside from metric moduli) turned on.  Our instability occurs because of \emph{global} properties of the handlebody, not due to any local properties of the metric.\footnote{A rather similar phenomenon was observed in higher dimensions in \cite{Belin:2014lea}, where hyperbolic black holes in AdS$_4$ were observed to undergo similar phase transitions even though the solutions were locally AdS.  As in the present case, the instability only arose because the hyperbolic black hole solutions differed globally from AdS$_4$. Thus modes of the scalar field which are not normally present (since they are non-normalizable in global AdS) suddenly become normalizable and lead to a genuine instability of the locally AdS solution.}   Although at first sight surprising, our results are a three dimensional version of a famous fact in two dimensions: the spectrum of the hyperbolic Laplacian on a Riemann surface depends not just on the local structure of the metric (which is always hyperbolic) but also on the moduli of the Riemann surface.  Indeed, this spectrum is the central object of interest in the study of arithmetic and quantum chaos, and many of our results are borrowed from this literature.  

Our results have important implications for entanglement entropies in two dimensional conformal field theories.  For any state in a two dimensional CFT, one can consider the reduced density matrix associated to a particular spatial region.  The R\'enyi entropies  ${1\over 1-n} \log \Tr \rho^n$ can then be used to characterize the spatial entanglement structure of this state.  When the spatial region is collection of intervals, the R\'enyi entropy is -- via the replica trick -- equal to the partition function of a CFT on a higher genus Riemann surface whose genus depends on $n$ (see e.g. \cite{Calabrese:2004eu}).   For example, the R\'enyi entropy for a pair of intervals in the vacuum state is equal to the partition function on a genus $g=n-1$ Riemann surface.  
The entanglement entropy is then computed by considering these R\'enyi entropies as an analytic function of $n$, and continuing to $n\to 1$.  
In this procedure one assumes that the entropies are analytic functions of $n$.  We have seen, however, that in holographic CFTs the R\'enyi entropies can undergo a phase transition as $n$ is varied, at some finite value of $n>1$.  Thus the replica method for computing entanglement (von Neumann) entropies must be treated with care.\footnote{Similar phenomena were observed for spherical entangling surfaces in higher dimensional holographic CFTs in \cite{Belin:2013dva, Belin:2014mva}, and for the three dimensional $O(N)$ model in \cite{Metlitski:2009iyg}. }  
For example, if we consider the R\'enyi entropies for a pair of intervals, two handlebodies will interchange dominance precisely at cross-ratio $x=1/2$ \cite{Headrick:2010zt}; this is also exactly where the Ryu-Takayanagi formula for entanglement entropy \cite{Ryu:2006bv} will undergo a phase transition.  Our results imply, however, that if the CFT has a sufficiently light operator then the R\'enyi entropies will undergo a phase transition at cross-ratio $x<1/2$.
For example, the $n=2$ R\'enyi entropy will undergo a phase transition if the theory has an operator with dimension $\Delta < \Delta_c= 0.189124\cdots$. 
We note that this instability occurs for values of $n$ which are strictly larger than one -- we do not expect a non-analyticity in a neighborhood of $n=1$.  It would be interesting to revisit the arguments of \cite{Faulkner:2013yia, Hartman:2013mia, Headrick:2010zt, Lewkowycz:2013nqa} in this context.  

Our results also make clear a sense in which higher genus CFT partition functions differ qualitatively from those on the sphere or torus.  The torus partition function, for example, was shown by Hartman, Keller and Stoica to take a universal form at large central charge, provided one assumes that the spectrum of light states (i.e. those with dimension less than the central charge) does not grow too quickly \cite{Hartman:2014oaa}.  This universal form is precisely that of a dual three dimensional theory of gravity which has a Hawking-Page transition between a thermal state and a BTZ black hole, and the sparseness condition is obeyed by any bulk local quantum field theory and even by string theories with string scale $\ell_\text{string} \lesssim \ell_\text{AdS}$.  At higher genus, however, we see that additional phase transitions are generic, and occur even for duals of local quantum field theories in the bulk.   Thus at higher genus there is no analogous ``universal partition function" at large central charge.

While our discussion will be entirely in the context of three-dimensional gravity, similar phenomena will also occur in higher dimensions. The most direct analogue is with locally AdS spacetimes, to which almost everything generalizes straightforwardly.  In particular, the critical dimension $\Delta_c$ for an instability can be shown to be equal to the Hausdorff dimension of an appropriate limit set, just as we will see below for the three dimensional case.\footnote{Indeed, one can directly reinterpret the results of \cite{Belin:2014lea} in this context.} This is slightly less natural than in three dimensions, because solutions to Einstein's equations need not be locally hyperbolic, so it is not clear when such geometries would dominate the path integral. More generally, the same mechanism of instability can apply, with global properties of the solution moving the critical mass above the na\"ive Breitenlohner-Freedman bound. Heuristically, scalars of negative mass squared can be stable because the reduction in action from the mass term in a finite region is compensated for by the positive contribution to the action from the gradient, required to match with the boundary conditions at infinity; it is important here that the volume of a region does not grow faster with size than its perimeter in negatively curved spaces. Without altering the local curvatures, nontrivial topology can upset this mechanism for stability by reducing the size of a region's boundary, and hence the gradient contribution to the action, for a given volume.



The discussion of the paper will be phrased in terms of the Euclidean solutions, but the results have interesting implications in Lorentzian signature. The relevant CFT states are defined by a Euclidean path integral on a Riemann surface with one or more boundaries, generalizing the familiar examples of the path integral on the disc preparing the vacuum state, and on the cylinder preparing the thermofield double state on two entangled copies of the CFT Hilbert space. This defines the state at $t=0$, which can be evolved in Lorentzian time.

To find the semiclassical bulk dual of these states, we must first find the Euclidean solution that dominates the path integral on the `Schottky double', the closed Riemann surface formed by gluing the surface to its mirror image along each of the boundaries, by construction producing a $\ZZ_2$-symmetric surface. The dominant solution is expected to respect this boundary time-reflection symmetry, so the bulk surface $\slice$ fixed by the reflection acts as an initial data surface for Lorentzian evolution, and the quantum state of the bulk fields is the Hartle-Hawking wavefunction on $\slice$. For one possible solution, the $t=0$ slice $\slice$ is conformal to the original Riemann surface, describing a single-exterior black hole with topology hidden behind a horizon for a single-boundary case, or a multi-boundary black hole with an exterior region for each boundary, all joined by a non-traversable wormhole. Even in pure gravity, there are several phases of the dominant bulk solution, so depending on the moduli the bulk state can also be disconnected copies of pure AdS (but with fields in a state different from the vacuum), or something else. For a more detailed review of these states, see \cite{Maxfield:2014kra,Maxfield:2016mwh,Skenderis:2009ju, Maloney:2015ina}.

Now, if there is a sufficiently relevant scalar operator in the CFT, there is an additional second-order phase transition to a dominant bulk solution with a nonzero classical value for the dual scalar field. This means that the initial data on $\slice$ includes some scalar field configuration, which will evolve in time. The fact that these states are not stationary will then be visible even for a classical observer outside any horizon. When the phase includes a black hole, the scalar outside the horizon will rapidly decay away, falling into the black hole. A more interesting time evolution occurs when the dual state includes copies of pure AdS, which may now include some scalar configuration. When the amplitude is small, as will be the case close to the transition, linearised evolution will suffice, with the field bouncing around periodically, but eventually nonlinearities will likely become important, with resonances between different modes. Perhaps the most likely evolution thereafter is a turbulent cascade to excite higher and higher frequency modes, with the solution nonetheless being regular for all time, as evidenced by numerical studies of a massless scalar interacting only gravitationally \cite{Bizon:2013xha}. This is different from the situation in higher dimensions, in which a black hole forms after finite time; this cannot occur in three dimensions, because there is a finite energy threshold between the vacuum and the lightest black hole\footnote{We would like to thank Benson Way for helpful comments on this aspect.}.


\hrulefill

In \cref{review} we will review briefly a few salient features of three dimensional gravity, as well as the necessary aspects of CFT on Riemann surfaces.  In \cref{CFTsec} we will give a CFT argument for the existence of an instability, inspired by recent results of Belin, Keller and Zadeh \cite{Belin:2017nze}.  The main idea is that a free bulk scalar field is dual to a generalized free field in the boundary CFT, and we can compute the contribution of such a field to the higher genus CFT partition function.  Using the asymptotic value of the OPE coefficients of multi-trace operators built from a generalized free field, along with higher genus conformal blocks in the appropriate regime, one can show that these contributions diverge when the field is sufficiently light, which signals the phase transition.  This argument allows us to bound the critical dimension of the scalar field; for example, for the genus two handlebody relevant for the computation of the third R\'enyi entropy of two intervals at cross-ratio $x=\frac{1}{2}$, we find $\Delta_c \geq 0.189121\cdots$.  

In \cref{bulkSec} we will turn to the bulk instability.  We will first review how the zero mode of the instability relates to various notions from the spectral theory of the Laplacian on a general bulk geometry $\bulk$. We then specialize to the main case of interest, for which $\bulk$ is a quotient of hyperbolic space $\HH_3$ such as a handlebody, and find that the critical dimension has a rather beautiful mathematical interpretation. The quotient is by a group of M\"obius maps, which has a limit set, a subset of the Riemann sphere (the boundary of Euclidean AdS$_3$). This limit set  has a finite Hausdorff dimension $\delta>0$, which is sometimes referred to as the fractal dimension of the limit set.  This Hausdorff dimension is precisely equal to the critical dimension of the scalar field, $\Delta_c= \delta$.  In other words, a scalar is unstable if and only if its dimension is less than the Hausdorff dimension of the limit set.  
The calculation in \cref{CFTsec} can therefore be regarded as a CFT estimate of this Hausdorff dimension, which provides explicit lower bounds on $\delta$.

In \cref{criticalDimSec} we turn to the explicit computation of the critical dimension, using an algorithm of McMullen for computing the Hausdorff dimension.  We will use the algorithm to compute the critical dimension analytically, finding the asymptotic behaviour of $\delta$ as the boundary of moduli space is approached, for the handlebody which dominates the partition function. We also describe what happens to the instability at large genus. We will also use the algorithm to efficiently compute the critical dimension numerically. For example, for the genus two surface described above, the Hausdorff dimension is 
$\delta = 0.189124\cdots$, close to our CFT bound.  We use the numerical data to provide plots of this critical dimension as a function of moduli, and as a function of genus.

\section{Review of higher genus partition functions in 3D gravity and 2D CFT}\label{review}

In this section we will review the description of higher genus Riemann surfaces, and the construction of solutions to three-dimensional gravity with such boundaries, which can be interpreted as saddle points for the higher genus partition function of a holographic CFT. In particular, we describe a class of symmetric surfaces that we will use as examples. We will also review the interpretation of these partition functions in terms of R\'enyi entropies.

\subsection{Moduli spaces and handlebodies} 

We are interested in studying holographic two-dimensional conformal field theories, dual to three dimensional AdS gravity in Euclidean signature, in particular on a Riemann surface $\bdry$ of genus $g\ge 2$. The partition function of the theory on such a surface, denoted $Z_g(\tau)$, will depend on the conformal structure of the surface $\bdry$.
%
%
%
Here $\tau$ is a collection of $3g-3$ complex coordinates which parameterize the moduli space
$\M_g$ of conformal structures on $\bdry$.
At genus one, $\tau$ can be identified with the usual torus modulus.
At higher genus there are various different coordinates which can be used to describe the moduli $\tau$, some of which we will now describe.\footnote{
Because of the conformal anomaly the partition function will in addition depend on a choice of metric within a given conformal class.   Thus $Z_g$  should not -- strictly speaking -- be regarded a function of $\tau$ alone.  This dependence, however, involves only the central charge and not any of the other dynamical data of the CFT (such as operator dimensions or structure constants), so will not be important for us here.  
We will therefore suppress this dependence and simply indicate the dependence on the conformal structure moduli $\tau$.}




For many purposes in CFT and gravity, the most convenient way to realize a Riemann surface $\bdry$ is as a quotient of the Riemann sphere $\CC^*$ by a \emph{Schottky group} $\Gamma$, $\bdry = \Omega(\Gamma)/\Gamma$.
Here $\Gamma$ is a discrete subgroup of $PSL(2,\CC)$, which acts on $\CC^*$ in the usual way by M\"obius transformations,   
and $\Omega(\Gamma)$ is the set of points on the Riemann sphere where this group `acts nicely'. More precisely, $\Omega(\Gamma)$ is the set of points $z\in\CC^*$ which have some neighborhood $U$ containing no other images of $z$ under the group: $\gamma\cdot z\in U$ for $\gamma\in\Gamma$ implies that $\gamma$ is the identity. Equivalently, if we define the limit set $\Lambda(\Gamma)$ to be the set of accumulation points of the action of $\Gamma$ on $\CC^*$ (a set about which we will have much more to say later), 
then $\Omega(\Gamma)$ is just the Riemann sphere with those points removed:
$\Omega(\Gamma)=\CC^* - \Lambda(\Gamma)$.  
More specifically, a Schottky group $\Gamma$ of genus $g$ is a subgroup of $PSL(2,\CC)$ that is freely generated by $g$ loxodromic\footnote{A loxodromic element $\gamma$ of $PSL(2,\CC)$ is one which is conjugate to $\begin{pmatrix} q & 0\\0 &q^{-1} \end{pmatrix}$ for some $q$ with $0<|q|<1$.} elements of $PSL(2,\CC)$, having as a fundamental domain the exterior of $2g$ closed curves (usually circles), such that each of the $g$ generators of $\Gamma$ maps one of these boundaries to another in pairs.
 Intuitively, to obtain a Schottky representation, we can cut the surface along $g$ disjoint closed loops such that it stays in one piece and becomes a sphere with $2g$ holes, flatten it onto the complex plane, and build the Schottky group from the M\"obius maps that glue the surface back together along its $g$ seams. A given Riemann surface can be written as a Schottky group in many different ways, depending on the choice of $g$ cycles to cut along. The presentation as a Schottky group is equivalent to the plumbing construction used in \cite{Cho:2017oxl}. A more detailed review of Schottky uniformization can be found in \cite{Krasnov:2000zq, Faulkner:2013yia, Barrella:2013wja}. A slightly different approach to calculations in the Schottky coordinates was used in \cite{Gaberdiel:2010jf}.

A rather different presentation of the Riemann surface $\bdry$ is as an algebraic curve.  In this case we represent $\bdry$ as the set of solutions to an equation such as
\begin{equation}\label{branched}
y^n = \prod_{k=1}^{N} \frac{z-u_k}{z-v_k}\;.    
\end{equation}
Here, $\bdry$ is a genus $g=(N-1)(n-1)$ surface, represented as $n$-fold branched cover over the Riemann sphere parameterized by the $z$-plane, with $2N$ branch points $(u_k, v_k)$.  As the resulting Riemann surface automatically possesses a $\ZZ_n$ symmetry (usually referred to as replica symmetry) where one permutes the $n$ sheets, one cannot describe a general point in moduli space $\M_g$ using this parameterization.  Instead, this equation describes only a $2N-3$ dimensional slice of moduli space, a family of surfaces with an enhanced ($\ZZ_n$) automorphism group.  Except in special cases one cannot map out the full moduli space this way.\footnote{One important special case is the genus two moduli space. Every genus two curve is hyperelliptic, so can be represented as a $2$-fold cover of the sphere branched over 6 points.}  The advantage of this approach is that the moduli of this surface are easy to describe -- they are the locations $(u_k, v_k)$ of the branch points. For physics, this presentation is natural for describing R\'enyi entropies (or certain correlation functions in orbifold theories), with the branch points $u_k$ and $v_k$ corresponding to the insertion points of twist and anti-twist operators.


Except at genus one, or in special cases with very high symmetry, it is not possible to find an explicit map between the moduli of the algebraic curve and Schottky groups, or to find out whether two Schottky groups represent the same surface, sliced in a different way. However, the problem of finding a Schottky group associated with a particular algebraic curve, sometimes called `Schottky uniformization', is equivalent to solving a monodromy problem, which we now briefly describe.

To do this, we begin by denoting the locations of branch points $(u_k, v_k)$ as $z_i$ ($i=1,\ldots,2N$). We would like to find the map $w(z)$ from the algebraic curve coordinate $z$ to the coordinate $w$ of the complex plane on which the Schottky group acts.  But $w(z)$ is not single-valued, because there are many possible values of $w$ related by elements of the Schottky group $\Gamma$. However, the Schwarzian derivative $T_c(z)=S(w)(z)=\left(\frac{w''}{w'}\right)'-\frac{1}{2}\left(\frac{w''}{w'}\right)^2$ is single-valued, since the Schottky group consists of M\"obius maps. If we take $T_c(z)$ as given, a simple calculation shows that solving $T_c(z)=S(w)(z)$ for $w$ is equivalent to solving the ordinary differential equation
\begin{equation}\label{ode}
        \psi''(z) + T_c(z) \psi(z) =0,
\end{equation}
with $w(z)=\frac{\psi_1(z)}{\psi_2(z)}$ being the ratio of two linearly independent solutions $\psi_{1,2}(z)$ to the ODE.

This is not much use if we know nothing about $T_c(z)$.  However, for the Riemann surface (\ref{branched}) $T_c(z)$ can be fixed up to a finite number of parameters, by using the fact that it is a meromorphic function of $z$ which transforms like a stress-tensor:
\begin{equation}
        T_c(z) = \sum_{i=1}^{2N} \left(\frac{1-n^{-2}}{4(z-z_i)^2}+\frac{\gamma_i}{z-z_i}\right)
\end{equation}
Here, we have assumed that the Schottky group respects the replica symmetry, so $T_c(z)$ is single-valued in $z$. The double poles are fixed by demanding smoothness in the $y$ coordinate of \cref{branched}, and the $\gamma_i$ are free parameters, called `accessory parameters'. It is also constrained by smoothness at infinity, which demands that $T_c(z)$ decays like $\frac{1}{z^4}$.  This imposes three constraints on the $\gamma_i$, leaving $2N-3$ free parameters.


It remains to fix these free parameters. To do this, note that if we go around a closed curve on the surface, a solution to the ODE will not usually come back to itself, but undergo monodromy, so the value of $w$ will change by a M\"obius map:

\begin{equation}
        \begin{pmatrix}
                \psi_1 \\ \psi_2
        \end{pmatrix} \longrightarrow
        \begin{pmatrix}
                a&b\\c&d
        \end{pmatrix}\begin{pmatrix}
                \psi_1 \\ \psi_2
        \end{pmatrix}, \quad\text{so that}\quad w\longrightarrow \frac{a w+b}{c w+d}
\end{equation}
The monodromies of the ODE form a representation of the fundamental group of the surface $\pi_1(\bdry)$ by M\"obius maps. But in the Schottky representation, not all the closed loops on the surface should take us to a different $w$, and a different copy of the fundamental domain for $\Gamma$: the $g$ special loops that bound the fundamental domain should come back to the same value of $w$, and so correspond to trivial monodromy of the ODE \cref{ode} (in fact, the monodromy matrix around these cycles is always minus the identity).


Imposing these trivial monodromy conditions is precisely enough to fix the $2N-3$ free parameters. 
Once these parameters are fixed, we may solve $\cref{ode}$ to find the monodromy around $g$ complementary cycles, which give the generators of the Schottky group $\Gamma$.
In \cref{exampleSurfaces} we go into more detail for a specific example, which we will subsequently use for analytic and numerical calculations.

This monodromy problem also appears in computations of the semiclassical limit of Virasoro conformal blocks \cite{zamolodchikov1987conformal}, reviewed in \cite{Harlow:2011ny,Hartman:2013mia}, and described in generality for higher genus blocks in \cite{Cardy:2017qhl}.

\subsection{Schottky representations, handlebodies, and gravity}

The Schottky representation has a very natural interpretation from the bulk point of view. To see this, note that the M\"obius maps acting on the Riemann sphere can be extended into a bulk hyperbolic space $\HH_3$, where they act as the orientation-preserving isometries. We can therefore extend the action of a Schottky group $\Gamma$ into this bulk, obtaining a quotient of hyperbolic space with $\bdry$ as its boundary, $\bulk=\HH^3/\Gamma$. Representing hyperbolic space in the upper half-space model, this can be understood as taking the circles that bound the fundamental domain of the Schottky group and extending them as hemispheres into the bulk, giving a fundamental domain with the hemispheres identified by the generators of $\Gamma$.

A CFT partition function on the surface $\bdry$ can be computed holographically as  the bulk gravity path integral over Euclidean geometries whose conformal boundary is $\bdry$.  Semiclassically, we just need to compute the action of a solution to the bulk equations of motion with  boundary $\bdry$, which will depend on the moduli of $\bdry$. There are an infinite number of bulk solutions, 
 which in general should include the contribution of matter fields, but a particularly simple class of solutions are those without matter fields turned on. Since pure gravity in three dimensions is locally trivial, Einstein's equations then imply that the bulk is locally $\HH_3$, which means that it must be a quotient of hyperbolic space. The Schottky group quotients therefore provide a large class of solutions to the bulk problem, which are conjectured to dominate the path integral in pure gravity.
 
  Topologically, the Schottky group quotients are handlebodies, obtained by `filling in' the surface $\bdry$ along a choice of $g$ cycles.
  These contractible cycles are precisely those we chose to cut the surface along to construct the Schottky group, or around which we imposed trivial monodromy. The Schottky group describes the remaining non-contractible cycles, in the sense that it is topologically interpreted as the fundamental group of the handlebody.
  
 Some geometric properties of the bulk can be read off easily from the Schottky group, in particular the lengths of closed geodesics. A closed loop is represented topologically as a conjugacy class in the fundamental group, or equivalently in $\Gamma$, and since the eigenvalues of $\gamma\in\Gamma$ are independent of the representative of the conjugacy class, the smaller eigenvalue $q_\gamma$ of $\gamma\in SL(2,\CC)$ ($0<|q_\gamma|<1$) is naturally associated with a closed curve. Writing $q_\gamma=e^{-\frac{1}{2}(\ell+i\vartheta)}$, $\ell$ is in fact the length of the closed geodesic, and $\theta$ is the amount the geodesic is twisted by (the angle a normal vector rotates by after parallel transport round the curve). Explicitly in terms of the trace, this length is
 \begin{equation}
        \ell_\gamma = \cosh^{-1}\left[\left|\frac{\Tr \gamma}{2}\right|^2+\left|\left(\frac{\Tr \gamma}{2}\right)^2-1\right|\right].
 \end{equation}
For more detailed review and discussion, focussing on the Lorentzian versions of these geometries, see \cite{Maxfield:2014kra}.

Because there are many possible Schottky groups corresponding to the same surface $\bdry$, we must decide which geometry gives the correct semiclassical bulk dual for given moduli of the boundary surface (even before considering bulk matter fields). The na\"ive answer to this, and the one that reproduces CFT expectations, is that the handlebody with least action dominates the path integral, so is the dual bulk. It was shown how to compute this action in \cite{Krasnov:2000zq}, from a particular higher-genus `Liouville action' \cite{zograf1988uniformization}, depending crucially on the IR cutoff imposed on the bulk, and hence on the choice of metric on the boundary surface within the given conformal class. In a metric appropriate for R\'enyi entropies, flat away from conical singularities at branch points $z_i$, and with a bulk preserving replica symmetry, the derivative of the action with respect to $z_i$ is proportional to the accessory parameter $\gamma_i$ \cite{Faulkner:2013yia}. In a constant curvature metric, for general surfaces, a numerical algorithm to compute the action was given in \cite{Maxfield:2016mwh}. As a heuristic, to choose the dominant saddle point, the $g$ shortest cycles of the surface should be filled in.

Given these tools, one can then attempt to construct the the higher genus partition function $Z_g(\tau)$ via a bulk path integral, as a sum over geometries.  The handlebodies described above give semi-classical saddle point contributions to this bulk path integral, and the full partition function should be given be a sum over these semi-classical saddles along with a set of loop corrections.  The loop corrections to these semi-classical contributions can be computed exactly at genus $g=1$ and perturbatively at higher genus  (see e.g. \cite{Maloney:2007ud, Headrick:2015gba}).  In pure gravity -- i.e. in theories with no degrees of freedom aside from the metric -- there is some hope that one could compute the higher genus partition function exactly \cite{Witten:2007kt}.  
We will be interested in more general theories, which contain scalar fields in addition to the metric.  In this case the theory has local bulk degrees of freedom, and there is little hope of an exact computation.  Nevertheless, the computations described above give contributions to the partition function of a holographic CFT which will be valid in the semi-classical (large $c$) limit.

\subsection{A $\ZZ_n$ symmetric family of genus $n-1$ surfaces}\label{exampleSurfaces}

We now illustrate this general discussion with an example, specifically a family of genus $n-1$ surfaces with an enhanced $\ZZ_n$ symmetry.
As an algebraic curve, this family of Riemann surfaces is given by 
\begin{equation}
y^n = \frac{z(z-1)}{z-x}\;.
\end{equation}
This is the $N=2$ case of (\ref{branched}), where  we have used $PSL(2,\CC)$ transformations to put $u_1=0$, $u_2=1$ and $v_2=\infty$. The remaining parameter is the cross-ratio $x$, which is the modulus of this family of Riemann surfaces.  In general $x$ can be any complex number, but for simplicity (and for the purposes of applications to R\'enyi entropies, described below) we will take it to be a real number between zero and one.

To find a Schottky group, or equivalently a bulk geometry, we can now solve the monodromy problem described above. Choosing to preserve the replica symmetry, the most general ansatz for the ODE \cref{ode} is
\begin{equation}
        T_c(z)=\frac{1-n^{-2}}{4}\left(\frac{1}{(z-x)^2}+\frac{1}{z^2}+\frac{1}{(z-1)^2}-\frac{2}{z (z-1)}\right)-\gamma\frac{x(1-x)}{z (z-1) (z-x)},
\end{equation}
where we have imposed the constraints (which are slightly different, because there is a branch point at infinity), leaving the single accessory parameter $\gamma$. To fix this parameter, we must first choose the cycles around which we impose trivial monodromy. For our purposes, it suffices to take the cycles surrounding $0$ and $x$; this gives a loop on the $z$ plane enclosing one zero and one pole of $y^n$, so remains on the same sheet of the branched cover, forming a closed loop on the surface. There are $n$ of these, one on each sheet, but only $n-1$ of them are in fact independent: the product (in the fundamental group) of the $n$ loops, described by a loop enclosing $z=0$ and $z=1$, then moving to the next sheet, and repeating $n$ times, is topologically trivial. Because we have imposed $\ZZ_n$ symmetry already on our ansatz for $T_c(z)$, imposing trivial monodromy on any one of the sheets is sufficient.

Having chosen the accessory parameters to trivialize the monodromy around these cycles, we would like to read off the Schottky group. To do this, it is convenient to take full advantage of the symmetry of the situation, using the automorphisms of the surface (which are preserved by the handlebody). In the language of the quotient, an (orientation-preserving) isometry of the bulk is represented by an additional element $\gamma\in PSL(2,\CC)$ (so it is an isometry on the covering space $\HH^3$) that commutes with the group $\Gamma$ ($\gamma\Gamma=\Gamma\gamma$, so it has a well-defined action on the quotient). Including some such elements, we can form an extended group $\hat{\Gamma}$, of which $\Gamma$ forms a normal subgroup. The largest possible $\hat{\Gamma}$, including all elements of $\gamma\in PSL(2,\CC)$ such that $\gamma\Gamma=\Gamma\gamma$ is the \emph{normalizer} $\mathcal{N}(\Gamma)$ of $\Gamma$, and the isometry group of the bulk is then $\operatorname{Isom}(\bulk)\simeq \mathcal{N}(\Gamma)/\Gamma$.

The most obvious extension providing an automorphism is the $\ZZ_n$ replica symmetry $R$, represented as the monodromy around a loop containing $0$. This comes back to a different sheet, so is not an element of $\Gamma$, but we can include it in $\hat{\Gamma}$ as an elliptic M\"obius map of order $n$.  From the point of view of the monodromy problem, these elements correspond to monodromy along curves that may not return to the same point on the surface, but go between some point and its image under the isometry. Near $z=0$, the independent solutions to the ODE \cref{ode} look like $\psi_\pm(z)\sim z^{\frac{n\pm 1}{2n}}$, with corrections forming a power series in $z$ and not affecting the monodromy around zero; choosing these as our basis $\psi_{\pm}$ of solutions ($w(z)=\frac{\psi_+(z)}{\psi_-(z)}$), the loop around zero, enacting the replica symmetry, acts on the $w$ coordinate as $R:w\mapsto e^{\frac{2\pi i}{n}}w$.

In fact, this family of surfaces automatically has more symmetry, containing an additional $\ZZ_2$ extending the $\ZZ_n$ to a dihedral group $D_{2n}$\footnote{Even further than this, these surfaces all have another additional $\ZZ_2$ commuting with this dihedral group, acting as $z\mapsto x\frac{z-1}{z-x}$, which is a hyperelliptic involution of $\mathcal{B}$. We will not make use of this extra symmetry, but in the parameterization used below, it can be included as $w\mapsto \frac{1}{w}$.}. From the $z$ coordinate, this can be understood as a map swapping $0$ with $1$ and $x$ with $\infty$, $z\mapsto \frac{x}{z}$, along with reversing the order of the sheets of the cover. It is straightforward to check that this leaves the ODE invariant, after transforming $\psi$ as a weight $-\frac{1}{2}$ field. In terms of the monodromy, this extra symmetry is enacted by taking the solutions $\psi_\pm(z)$, following the solution from $0$ to $\infty$, and reading off the coefficients of $z^{\frac{n\pm 1}{2n}}$ in these solutions near $\infty$, giving some order two M\"obius map $S$.

In practice, except for the special case $n=2$, finding $S$ requires doing the calculation numerically, but we can deduce a lot about it, reducing the unknown parameters from the three numbers specifying a general M\"obius map, to just one. Firstly, note that doing $S$ twice corresponds to going round a loop with trivial topology, which implies that $S$ is order two, $S^2=\mathbb{1}$, which means it is specified by its fixed points. Secondly, without altering the form of $R$, we can change coordinates by rescaling and rotating in the $w$ plane, and use this to remove one other parameter of freedom. We will use this freedom to set the product of the fixed points of $S$ to be unity, which fixes $S$ to act as $S:w\mapsto \frac{w-\zeta}{\zeta w-1}$ for some (in general complex) $\zeta$.

Now, the extended group $\hat{\Gamma}$ is generated by just $R$ and $S$ (in fact, it is the free group generated by those elements with the only relations being those given by the orders of the elements: $\hat{\Gamma}=\langle R,S | R^n=S^2=\mathbb{1}\rangle$), so we are interested in the one-parameter family of groups generated by one M\"obius map of order $2$, and one of order $n$. The actual Schottky group $\Gamma$ appears as a normal subgroup of this, generated by the loops that actually return to the same point on the surface (but are still non-contractible in the bulk), requiring an even number of $S$ generators to appear, and also for the sheets to map back to themselves, rather than being permuted. The first of these is the element $\gamma_1 = SRSR$, taking a loop round zero by $R$, then going from $0$ to $\infty$ by $S$, then a loop round to infinity by $R$ again, and finally back to the starting point at zero, creating a closed loop surrounding zero and infinity, or equivalently $x$ and $1$. The remaining generators are similar, but starting on a different sheet, achieved by conjugating with $R$: $\gamma_k = R^{1-k}SRSR^k$. There are $n$ of these, but they are not all independent, since $\gamma_1\gamma_2\cdots\gamma_n=\mathbb{1}$. Any $n-1$ of these (a number equal to the genus) generate $\Gamma$. To relate this to the general discussion of symmetries above, the group of isometries $\hat{\Gamma}/\Gamma$ described by this extension is the dihedral group of order $2n$, since modding out by $\Gamma$ is equivalent to imposing the additional relation $\gamma_1=SRSR=\mathbb{1}$, giving the presentation $\hat{\Gamma}/\Gamma=\langle R,S | R^n=S^2=SRSR=\mathbb{1}\rangle\equiv D_{2n}$.

This prescribes the family of Schottky groups we are interested in, parameterized by $\zeta=\cos\theta$, though it is important to note that this only describes a Schottky group when $\zeta$ is sufficiently close to one (or $\theta$ close to zero). An alternative, more geometric parameterization is by $q$, defined as the smaller eigenvalue of $\gamma_1=SRSR\in SL(2,\CC)$ (noting that this is independent of the sign chosen for the matrix representatives of $S$ and $R$), defined as above so that $q=e^{-\frac{1}{2}(\ell+i\vartheta)}$ gives the length and twist of a curve in the bulk geometry. In the case $n=2$, $q=\tan^2\frac{\theta}{2}$ is the usual elliptic nome of the boundary torus, lying in the punctured open unit disc, though for larger $n$ it must be contained in a strictly smaller region.

 In practice, to map from any given $x$ to find the corresponding value of $\zeta$ (or $\theta$ or $q$), it is sufficient to compute the trace of any of the $\gamma_k$ (all are equal), from the trace of the monodromy of any loop containing $x$ and $1$. It is also possible to solve the monodromy problem perturbatively in small cross-ratio, as described in \cite{Barrella:2013wja}, and in our parametrisation, the result to leading order is $\theta=\frac{\sqrt{x}}{n}(1 + O(x))$. Real $x$ between $0$ and $1$ corresponds to real $\theta$, or $0<\zeta<1$, or $0<q<1$. Real negative $x$ (or equivalently $x>1$) also results in a Fuchsian group $\Gamma$, and corresponds to $\zeta>1$ (but bounded by $\sec\frac{\pi}{n}$ so the group is Schottky).

We can also find different handlebodies for the same surface by trivializing monodromy around some different cycle, but the only possibilities preserving the replica symmetry are much the same, the most obvious being to take the loop to surround $x$ and $1$ rather than $0$ and $x$. There is a phase transition between the two corresponding handlebodies at $\Re(x)=\frac{1}{2}$ \cite{Headrick:2010zt}.

\subsection{Relationship with R\'enyi entropies}

 The bulk computation of higher genus partition functions can be applied to the computation of R\'enyi entropies in holographic CFTs (see \cite{Calabrese:2009qy} for a review).  For a density matrix $\rho$, the $n$th R\'enyi entropy is defined as 
\begin{equation}
        S_n = \frac{1}{1-n} \log \Tr \rho^n.
\end{equation}
In the $n\to1$ limit this becomes the von Neumann entropy $S = -\Tr(\rho\log\rho)$.  In order to probe the spatial entanglement structure of the theory, we can take $\rho$ to be the reduced density matrix for a spatial region $A$ in the vacuum state.  Then $\rho$ is defined by a Euclidean path integral on the sphere (with cuts introduced at A),  and the R\'enyi entropy may be computed by gluing $n$ copies of this sphere together along these cuts.  Explicitly, we have 
\begin{equation}
S_n = \frac{1}{1-n} \log \frac{Z_n}{Z_1^n}.
\end{equation}
Here $Z_n$ is the partition function on a manifold $M_n$, which is the $n$-fold branched cover defined by gluing $n$ copies of the original spacetime manifold along $A$, and the normalization constant $Z_1$ is the sphere partition function.  If $A$ consists of $N$ disjoint intervals, this is precisely the $n-$fold cover of the sphere branched over $2N$ points (the endpoints) described above. 

We will focus on the case where $A$ consists of two disjoint intervals $[u_1,v_1]$ and $[u_2,v_2]$.  Then the conformal structure of $M_n$ is completely determined by the cross ratio
\begin{equation}
x = \frac{(u_1-v_1) (u_2-v_2)}{(u_1-u_2) (v_1-v_2)}.
\end{equation}
As this cross-ratio is varied, we sweep out a one (real) dimensional slice of the moduli space $\M_{n-1}$ of genus $n-1$ dimensional Riemann surfaces.
This is precisely the case described in the previous subsection.
There are two handlebodies which compute the pure-gravity contribution to the R\'enyi entropies, which exchange dominance at the point $x=1/2$.   When $n=2$ these saddles are precisely the thermal AdS and Euclidean BTZ black hole solutions, and the phase transition at $x=1/2$ is the usual Hawking-Page phase transition.  The scalar instabilities we describe in this paper will occur for $n>2$ when the theory has an operator of dimension $\Delta$ which is sufficiently light.  In particular, we will find that there are two new phase transitions as $x$ is varied, one at $x=x_c(\Delta) <1/2$ and one at $x=1-x_c(\Delta)$, where these two handlebodies will become unstable to the formation of scalar hair.



\section{The phase transition from CFT}\label{CFTsec}

In this section we will make a CFT argument for the instability, by considering the contribution of a generalized free field -- the boundary avatar of a free bulk scalar field -- to the higher genus partition function of a CFT with large central charge. The result is that if the corresponding operator is sufficiently light, then the generalized free partition function will diverge somewhere in the interior of moduli space.  This signals that the free approximation has broken down, so interactions become important, and the partition function will undergo a phase transition. We give an analytic lower bound on the critical dimension $\Delta_c$ in terms of the Schottky moduli of the surface.  As the field becomes lighter, the phase transition will occur closer to the boundary of moduli space; in particular, for a sufficiently light field the corresponding bulk phase transition will occur before the usual ``Hawking-Page" transition where (locally Einstein) bulk saddles are interchanged.  In the bulk, this would be interpreted as the condensation of a bulk scalar field in a handlebody background.  The discussion in this section is a refinement of the arguments presented in \cite{Belin:2017nze}. 

\subsection{The partition function and conformal blocks}

A higher genus partition function can, at least in principle, be computed from the basic dynamical data of the CFT, namely the spectrum of dimensions and spins $(\Delta_i, s_i)$ of primary operators, along with their three-point coefficients $C_{ijk}$. To do this, we can insert a complete set of states on a handle of the surface to reduce the computation to sum over two-point functions on a surface one genus lower, and repeat this (along with use of the OPE) until the computation has been reduced to three-point functions on spheres. A complete decomposition like this can be understood by cutting up the surface into pairs of pants: any genus $g\geq 2$ surface can be decomposed (in many ways) into $2(g-1)$ pairs of pants, joined along a total of $3(g-1)$ cuffs. Along each of these cuffs, we can insert a complete set of states and, by the state-operator correspondence, the amplitude between three states defined by the path integral on the pair of pants is determined by a three-point coefficient.

Along with inserting complete sets of states in this way, we can use the fact that the states are arranged in multiplets of the Virasoro algebra, by summing up all contributions from a given multiplet appearing in the sums over states. The resulting object, collecting the contributions from a given primary on each of the $3(g-1)$ cuffs along with all their descendants, is a higher genus conformal block $\mathcal{F}$. This is determined by kinematics alone, depending only on the scaling dimensions and spins of the primaries chosen, the central charge, and the moduli of the surface\footnote{The blocks also factorize into the product of holomorphic and antiholomorphic blocks, though we will not explicitly use this fact here.}. In the partition function this conformal block will be multiplied by the product of $2(g-1)$ OPE coefficients corresponding to the primaries.
 
  Summing over all possible choices of primaries, the result is a general expression for the partition function of the form
\begin{equation}\label{frodo}
Z_g(\tau) = \sum_{\left\{i\right\}}  C^{2(g-1)}_{\{i\}} \mathcal{F} (\left\{\Delta_i\right\},c;\tau)
\end{equation}
Here the sum is over all choices of the $3(g-1)$ primary operators, and $C^{2(g-1)}_{\{i\}}$ denotes the product of $2(g-1)$ OPE coefficients corresponding to the primary operators propagating down the legs of each pair of pants. This expression may look very different for different pair-of-pants decompositions of the surface, but the result must be equal whichever decompositions is chosen; this is the statement of higher-genus crossing symmetry, which can be exploited to constrain CFT data \cite{Cardy:2017qhl,Cho:2017fzo,Keller:2017iql}.

\begin{figure}
	\centering
	\includegraphics[width=.6\textwidth]{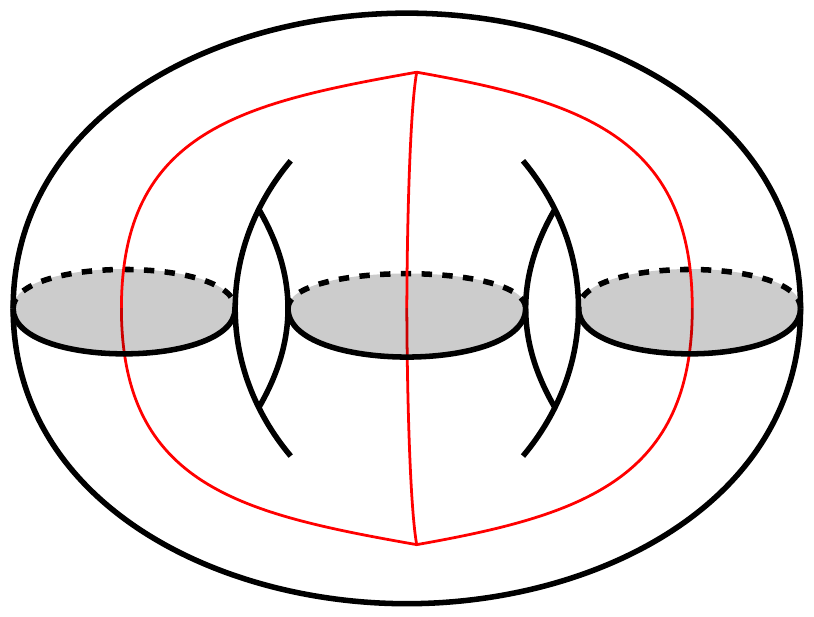}
	\caption{A genus 2 surface, which is cut into two pairs of pants glued together along the three black circles.  Along each of the three circles we can insert a projection onto the descendants of a primary of dimension $\Delta_i$ ($i=1,2,3$) to obtain the block $\mathcal{F}(\{\Delta_i\},c;\tau)$.  The dual handlebody is found by `filling in' the surface, as indicated by the shaded disks.  The block $\mathcal{F}(\{\Delta_i\},c;\tau)$ can be computed in the bulk in a semi-classical approximation, valid in the limit $1\ll\Delta_i\ll c$, by computing the action of the network of bulk geodesics indicated in red.\label{blockfig}}
\end{figure}

In the case $g=2$, there are two possible distinct types of decomposition into pairs of pants, depending on whether we choose to insert a complete set of states on a cycle dividing the surface into two pieces. Assuming we do not, the decomposition looks like two pairs of pants joined to one another along each of their three cuffs, as illustrated in \cref{blockfig}, and $C^{2(g-1)}_{\{i\}}$ is just $C_{ijk}^2$, where $i,j,k$ denote the primaries chosen on each of these seams.

In general, it is rather difficult to compute \eqref{frodo} explicitly, and even the conformal blocks cannot be calculated exactly in closed form. It is possible to calculate perturbatively in moduli of the surface, or using recursion relations exploiting the structure of degenerate representations \cite{Cho:2017fzo,Cho:2017oxl}, or in various semiclassical limits \cite{Cardy:2017qhl,Kraus:2017ezw}. In particular, one needs to choose a conformal frame, and to account carefully for the way in which the surface is glued together from its constituents.  Fortunately we will not need to work with this expression in generality, only requiring the blocks in a particular `semiclassical global' limit.

\subsection{Semiclassical global limit of higher genus blocks}

We will require the blocks in a limit of large central charge, where the dimensions of exchanged operators are large also large (with ratios between different $\Delta_i$ fixed in the limit), but small compared to $c$: $1\ll\Delta_i\ll c$. This limit has a dual holographic description in terms of semiclassical gravity coupled to particles in a probe limit, for which only the global conformal $\mathfrak{sl}(2)$ subalgebra of the Virasoro algebra is important; hence the name `semiclassical global' blocks.

In this limit the blocks simplify, becoming
\begin{equation}
        \mathcal{F} (\left\{\Delta_i\right\},c;\tau) = e^{-c S_0(\tau) -\Delta S_1\left(\left\{\Delta_i/\Delta_j \right\};\tau\right) + O(1,\Delta^2/c)},
\end{equation}
The functions $S_0$ and $S_1$, depend on the moduli and (in the case of $S_1$) on the ratios of conformal dimensions, and have semi-classical gravity interpretations which we will describe below. The fact that the blocks exponentiate in this limit large $c$ limit is most well known in the case of four-point functions \cite{Zamolodchikov:1985ie}, but has not been rigorously proven.  It is, however, physically well-motivated, for example by considering a semiclassical limit of Liouville theory \cite{Harlow:2011ny}.

First, $S_0$ can be interpreted as the semiclassical vacuum block, i.e. the block for which all operators are taken to be the identity.  It is equal to the on-shell action of the handlebody where each of the cycles in the pair-of-pants decomposition are chosen to be contractible.\footnote{This only depends on the choice of $g$ cycles, not the full decomposition, which is consistent because of the fusion rules of the identity: choosing the identity module along $g$ cycles is enough to imply that the identity must be present in the other $2g-3$.} This depends on a choice of conformal frame (the conformal anomaly precisely takes the form of a shift in $S_0$), and in general can only be computed exactly by numerics \cite{Hartman:2013mia,Faulkner:2013yia,Maxfield:2016mwh}. The frame-independent information contained in $S_0$ is the difference between its value in different channels, and this data is required to impose crossing symmetry. Luckily, for our purposes we will only need to express the partition function in one channel, so we can entirely disregard this piece.

For us, the more important contribution is $S_1$, encoding the dependence on the dimensions. The important point is that in the limit $c\to\infty$ with fixed $\Delta_i$ the contribution of the Virasoro descendants is unimportant after we factor out the vacuum block contribution $S_0$.  So the block reduces to a `global' block, where only the $L_{-1}$ descendants are kept \cite{Cho:2017oxl}\footnote{What constitutes a global descendant is a little ambiguous for higher-genus surfaces, since it is not invariant under general conformal transformations. The statement here requires a Schottky, or plumbing frame, for which all transition maps are M\"obius maps.}. The gravitational interpretation is clear: as $c\to\infty$, the backreaction from the matter and the loop corrections from the graviton can be ignored, and we need only the classical background action.

This global block is still tricky to compute at higher genus, but if we further assume that the internal dimensions $\Delta_i$ are large (but still much smaller than $c$), it simplifies to a `semiclassical global' block. This can be determined by considering a network of geodesics in the handlebody spacetime, determined by the pair-of-pants decomposition. Specifically, for each pair of pants, assign a trivalent vertex in the bulk, and join these vertices by a geodesic for each seam joining the pairs of pants. This geodesic is interpreted as the worldline of a particle of mass $m_i\sim \Delta_i$, determined by the dimension of the primary operator assigned to the corresponding cuff of the pants decomposition, and is assigned an action $\Delta_i \ell_i$, where $\ell_i$ is the length of that geodesic segment. We finally must specify the bulk locations of the vertices; these are chosen to minimize the total particle action $\sum_i \Delta_i \ell_i$. The value of $e^{-\min{\sum_i \Delta_i \ell_i}}$ at this minimum obeys the semiclassical limit of the Casimir equations for the global conformal blocks, as shown in \cite{Kraus:2017ezw}, so reproduces the correct moduli dependence of the blocks.\footnote{For some values of the dimensions and moduli, the action can be minimized when one of more of the geodesics shrink down to zero size, in which case the block is given instead by some complexified saddle point. From the bulk point of view, this happens when the leading order amplitude in the large $\Delta$ limit comes from double-trace contributions, rather than the case we would like to consider, where these are exponentially suppressed in $\Delta$ relative to the single traces \cite{Maxfield:2017rkn}.} This limit can be used as a starting point for a systematic perturbative expansion for the blocks, developed in terms of worldline quantum field theory coupled to gravity in \cite{Maxfield:2017rkn}.

The only thing that remains to fix is the overall normalization. The geodesic network prescription comes with an unambiguous normalization, but rather than being the canonical one, where we multiply by the appropriate OPE coefficients to find the contribution to the partition function, it comes with some nontrivial OPE coefficients $\hat{C}(\Delta_i)$ built in, depending on the dimensions of the operators meeting at each vertex. To compute these, consider taking the pinching limit in which all cuffs of the pairs of pants become small, suppressing the descendants so only the product of primary three-point functions remains. The function $\hat{C}(\Delta_i)$ can therefore be computed by using a geodesic approximation to a three-point function, with three geodesics going from the boundary of AdS and meeting at a trivalent vertex \cite{Chang:2016ftb}:
\begin{align}
        &\hat{C}(\Delta_i)=e^{\mathcal{P}(\Delta_i)},\text{ where} \\
        \mathcal{P}(\Delta_i)&=\frac{1}{2}\Delta_1\log\left[\frac{(\Delta_1+\Delta_2-\Delta_3)(\Delta_1+\Delta_3-\Delta_2)}{\Delta_2+\Delta_3-\Delta_1}\right] + (\text{2 permutations}) \nonumber \\
        &\quad\textstyle +\frac{1}{2}\left(\sum_i \Delta_i\right)\left(\log \sum_i\Delta_i-\log 4\right)-\sum_i \Delta_i\log\Delta_i \nonumber
\end{align}
Alternatively, the same result can be obtained from an appropriate limit of the DOZZ formula \cite{Dorn:1994xn,Zamolodchikov:1995aa}. The function $\mathcal{P}$ is homogeneous of degree one in the dimensions, so gives a contribution scaling linearly with dimension in the exponential, as required. We will make particular use of the special case where all dimensions $\Delta_i$ are equal to $\Delta_p$, for which $\mathcal{P}=-\frac{3}{2}\log(\frac{4}{3})\Delta_p$.

In the end, this gives the expressions for the blocks that we will use, applying in the limit $1\ll \Delta_i\ll c$:
\begin{equation}
        \mathcal{F}(\Delta_i,c)\sim \exp\left[-c S_0-\min\left\{\sum_\text{edges}\Delta_i\ell_i\right\}-\sum_\text{vertices}\mathcal{P}(\Delta_i)\right]
\end{equation}
We will apply this result in the specific case of the $\ZZ_3$ symmetric, genus 2 handlebodies, with time-reflection symmetry, corresponding to the $n=2$ version of the example in \cref{exampleSurfaces}, with $x$ (or $\theta$) real. The relevant geodesic network for the channel of interest is shown in red, in \cref{blockfig}. Furthermore we will take the dimensions of the three internal operators to be equal, $\Delta_1=\Delta_2=\Delta_3=\Delta_p$. In this case, finding the location of the vertices is straightforward, since they are fixed completely by symmetry, absolving us of the need to solve the minimization problem. It is now a simple exercise in hyperbolic geometry to work out the length of the geodesics connecting the two vertices, finding $\ell=\log\left(\cot^2 \frac{\theta}{4}\right)$. This, along with $\mathcal{P}=-\frac{3}{2}\log(\frac{4}{3})\Delta_p$ as computed above, gives the result we will need for the block:
\begin{equation}\label{blocksresult}
         \mathcal{F}_{g=2}(\Delta_p,c)\sim e^{-c S_0} \left(\frac{4}{3}\tan^2\frac{\theta}{4}\right)^{3\Delta_p}
\end{equation}

The intuition behind the derivation of this expression relies on the operators in the internal channels being single trace operators, corresponding to single particle states, in a theory with semiclassical gravity dual. But because the blocks are kinematic objects, these restrictions are not required to apply the formula. We will use it in the case where the internal operators are highly composite multi-traces built from a primary of small dimension, for which the intuition behind the semiclassical blocks certainly does not hold.

\subsection{Applying the blocks to generalized free fields}

Consider a scalar $\op$ of dimension $\Delta$, dual to a weakly interacting bulk field. As long as these interactions are unimportant, we can treat $\op$ as a generalized free field, which means that we can sensibly talk about composite `multi-trace' operators built from products of $\op$ and derivatives, $:\!\!\partial^\#\op \cdots \partial^\#\op\!\!:$. In the generalized free approximation, the dimensions of these products simply add, and they have vanishing connected correlation functions, so the correlators can be computed by Wick contractions.


Now, let's try to compute the genus two partition function using the conformal block decomposition, accounting for such a free bulk scalar field. It is a slightly tricky prospect accounting for all the possible multi-trace exchanges, so we will make a slightly crude approximation, taking the contribution only of primary operators $:\!\op^K\!\!:$ without derivatives, of dimension $K \Delta$, and also taking the same operator to propagate in all three legs.
This gives us a lower bound for the partition function, since the OPE coefficients are real and the blocks are positive\footnote{To prove this, note that when $x$ is real, the surface can be constructed by gluing a pair of pants directly to a reflected version of itself. The path integral on the pair of pants defines a state on three copies of the CFT, and the block (times OPE coefficients) is the expectation value of a projection (a positive operator) in this state, which is positive by unitarity.}:
\begin{equation}
        Z_\text{GF} \geq \sum_K C_{KKK}^2 \mathcal{F}_{g=2}(K\Delta,c)
\end{equation}
  The OPE coefficients appearing can be computed from the combinatorics of the Wick contractions \cite{Belin:2017nze}, and for three identical operators $:\!\op^K\!\!:$, in the limit of large $K$, the result is
\begin{equation}
        C_{KKK} \sim 2^{3K/2}.
\end{equation}
Putting this together with \cref{blocksresult} giving the blocks in the appropriate limit, the terms in the sum for large $K$ look like
\begin{equation}
        C_{KKK}^2 \mathcal{F}_{g=2}(K\Delta,c) \sim 2^{3K}\left(\frac{4}{3}\tan^2\frac{\theta}{4}\right)^{3K\Delta}.
\end{equation}
But now, if $\Delta$ is sufficiently small, these terms grow exponentially, causing the partition function to diverge! We can therefore put a bound on the critical dimension $\Delta_c$ at which $Z_\text{GF}(\Delta)$ diverges:
\begin{equation}
        \Delta_c \geq \frac{\log2}{\log\left(\frac{3}{4}\cot^2\frac{\theta}{4}\right)}.
\end{equation}
We do not expect this to be exact, since we have dropped the contribution of so many operators, but we will see later that it becomes asymptotically equal to the correct value at small $\theta$, corresponding to small $x$. For cross-ratio $x=\frac{1}{2}$, numerically solving the monodromy problem described around \cref{ode}, we find the corresponding value $\theta=.55128$. This gives the bound $\Delta_c\geq0.189219$ on the critical dimension (accurate to the number of quoted decimal places), in agreement with the analysis of \cite{Belin:2017nze}.

A partition function should be well-defined for any surface, so it may seem puzzling to get a divergent answer. The resolution is that the partition function is not truly divergent, but our approximations on the spectrum and OPE coefficients do not apply when $K$ is parametrically large. Even if we do not give a potential to the bulk field, it interacts through gravity, so the approximation of computing OPE coefficients of multi-trace operators by Wick contractions will cease to apply when $K$ is of order $\sqrt{c}$, 
though it could break down sooner if other interactions become important at a lower energy scale. When we pass the critical dimension, the sum over blocks will shift from being dominated by the vacuum, to being dominated by the multi-particle states at a scale set by the interactions. This signals a second order phase transition, which we will explain from the bulk as condensation of the scalar field.

\section{The bulk instability}\label{bulkSec}

We have argued in Section 3 that the contribution of a free scalar to the genus two partition function will diverge for sufficiently small conformal dimension, $\Delta<\Delta_c$. This divergence comes from the contribution of multi-trace states which are dual in the bulk to states with large particle number.  It is therefore natural to expect that the divergence signals an instability where the bulk scalar field condenses to form a new solution with a nonzero classical value. This implies the existence of a second-order phase transition, below which the semiclassical bulk path integral is dominated by a new classical solution of the (nonlinear) bulk equations of motion: a `hairy handlebody'.

The new classical solution will depend on the details of the theory, and in particular the interactions of the bulk field. These interactions give anomalous dimensions and couplings to the multi-trace operators in the theory, which become important above the scale of the interactions.\footnote{This scale could be the Planck scale for a free scalar minimally coupled to gravity, the AdS scale for a strongly coupled bulk field, or some intermediate energy scale such as the string scale.}  In particular, they will modify the asymptotic behaviour of the sum described in section 3 in such a way as to cure the divergence. The result is that the partition function will have some non-universal contribution at the interaction scale of the bulk field.

While the full nonlinear solution depends on details of the theory, the onset of the instability does not, and is sensitive only to the background geometry and the mass of the scalar field. It is characterized by the appearance at $\Delta=\Delta_c$ of a zero mode, a nonzero solution of the linearized bulk wave equation with source-free boundary conditions, which corresponds to a flat direction in the path integral. In this section, we will show that such a zero mode exists in quite general circumstances, and characterize the critical dimension $\Delta_c$ in terms of the bulk geometry.

\subsection{The zero mode and spectral theory}

In a $d$-dimensional holographic CFT, a single-trace scalar operator $\op$ of dimension $\Delta$ is dual to a bulk scalar field $\phi$ of mass $m^2=\Delta(\Delta-d)$. The linearized bulk equation of motion $(\nabla^2-m^2)\phi=0$ has two linearly independent solutions with different asymptotic behaviour near the boundary:
\begin{equation}\label{boundaryExpansion}
        \phi(x,z) \sim J(x)z^{d-\Delta} + \langle\op(x)\rangle z^\Delta
\end{equation}
Here $J(x)$ is a source for the operator $\op$ in the CFT, which is fixed as a boundary condition, and the expectation value  $\langle\op(x)\rangle$, in most circumstances determined uniquely by the boundary condition $J$ and regularity, describes the `response' of the scalar field in the presence of the linearized source $J$. We will be most interested in relevant operators $\Delta<d$, corresponding to masses which are na\"ively tachyonic, but above the Breitenlohner-Freedman (BF) bound, $-\frac{d^2}{4}< m^2<0$. In particular, we recall that for $-\frac{d^2}{4}<m^2<-\frac{d^2}{4}+1$, there are two possible choices of boundary condition for the scalar field $\phi$ with unitary duals \cite{Klebanov:1999tb}. 
These two different boundary conditions correspond simply to a choice of which of the two boundary behaviours in \eqref{boundaryExpansion} one chooses to view as a source, and which as a response. 

This linearized Klein-Gordon equation suffices to find leading order correlation functions (ignoring backreaction and other interactions), not just on pure AdS, but any asymptotically AdS geometry $\bulk$ obtained altering the boundary geometry or sourcing other fields, as long as $\phi=0$ on the background.\footnote{In general, $\phi$ could have couplings to curvature or other nonzero fields which modify this linearized equation, but we will be interested primarily in geometries which are locally AdS. Thus all of these couplings may be incorporated into an effective bulk mass of the scalar field.} This corner of AdS/CFT therefore reduces to the theory of the Laplacian on the manifold $\bulk$. Even for geometries that are locally AdS, this spectral theory can be rich and interesting, and we will import some ideas and results from the mathematics literature and explore the physical consequences.


Our key result is that, even in the absence of a source for the operator $\mathcal{O}$, it is possible for $\phi$ to spontaneously acquire a nonzero classical expectation value. The second-order transition to this behaviour occurs when there is a nonzero solution of the bulk wave equation 
\begin{equation}
\left(\nabla^2-\Delta(\Delta-d)\right)\phi=0
\end{equation} with vanishing source $J=0$.  In other words, as we vary the bulk solution $\bulk$ (or the dimension $\Delta$), the solution will become unstable when there is an eigenfunction of the Laplacian with boundary condition $J=0$ and eigenvalue $\Delta(\Delta-d)$. For a given geometry $\bulk$, we call the largest dimension for which this occurs the critical dimension $\Delta_c$. Reducing $\Delta$ further, this eigenfunction becomes a mode which decreases the action of the solution, so a given geometry is unstable to condensation of a scalar with $\Delta<\Delta_c$.

It is easy to see that the instability cannot happen if $\Delta>d$.
In particular, for a scalar field with $(\nabla^2-m^2)\phi=0$ we can use the standard argument for negativity of the Laplacian: 
\begin{equation}
        0\leq \int_\bulk (\nabla\phi)^2 = -\int_\bulk \phi \nabla^2\phi =-m^2\int_\bulk \phi^2 \implies m^2<0
\end{equation}
Here we have integrated by parts, and used the fact that the fast fall-off conditions ($J=0$ and $\Delta>\frac{d}{2}$) imply that all integrals converge and boundary terms vanish.
This instability is therefore ruled out for an irrelevant operator, but not immediately excluded for relevant operators, for which the boundary terms do not automatically vanish. We will see that such instabilities do occur, and are in fact quite generic.

Before giving our first characterization of the onset of the instability in terms of spectral theory, we should first clarify some mathematical terminology.  We are seeking a solution $\phi$ of the equation
$\nabla^2\phi = \Delta(\Delta-d)\phi$, with boundary conditions $J=0$.
In the mathematics literature, this would be called an eigenfunction of the Laplacian only if $\Delta>\frac{d}{2}$, since in this case the boundary condition $J=0$ is equivalent to demanding square-integrability of the eigenfunction: $\phi\in L^2(\bulk)$. We will also be interested in the alternate quantization of the scalar field, corresponding to operators in the range $0<\Delta<\frac{d}{2}$, where the boundary condition is imposed on the slowly-decaying mode. The dimension $\Delta$ then corresponds to a \emph{resonance} of the Laplacian, which is defined as a pole of the resolvent operator $R_\Delta =(\Delta(\Delta-d)-\nabla^2)^{-1}$, analytically continued in $\Delta$. The resolvent is essentially the bulk Green's function (bulk-to-bulk propagator) $G_\Delta$ on $\bulk$; to compute the action of the resolvent on a function, integrate it against $G_\Delta$, which satisfies the bulk wave equation with delta-function source:
\begin{equation}
        R_\Delta[\phi](y) = \int_\bulk d^{d+1}y' G_\Delta(y,y') \phi(y'),\quad (\Delta(\Delta-d)-\nabla_y^2)G_\Delta(y,y')=\delta(y,y')
\end{equation}
The critical dimension $\Delta_c$ will therefore show up as a pole in the bulk Green's function $G_\Delta$. 

Another way to characterize the critical dimension $\Delta_c$, is to note that the determinant of the resolvent $\det (\Delta(\Delta-d)-\nabla^2)^{-1}$ is precisely the square of the one-loop partition function of $\phi$.  One can therefore find $\Delta_c$ by looking for a divergence in the one-loop contribution of a scalar field $\phi$ on the background $\M$. 

As a final characterisation of $\Delta_c$, we can consider the linear response problem of turning on some small source $J(x)$, solving the bulk wave equation with the corresponding boundary condition, and reading off the response $\langle\op\rangle_J$. At generic values of $\Delta$, this problem will have a unique solution, so defines a linear map $S_\Delta:J\mapsto \langle\op\rangle_J$ between functions on the boundary $\bdry$, known in the mathematics literature as the scattering matrix. If we tune $\Delta$ to the critical dimension $\Delta_c$, however, there is an ambiguity, as we can always add a multiple of the zero mode to the solution. The zero mode therefore also shows up as a pole of $S_\Delta$, a \emph{scattering pole}.\footnote{Scattering poles do not coincide with the resonances for two reasons. The first is that there are also zeros of the scattering matrix, corresponding to solutions with a source but zero response, which may cancel a pole, giving a resonance without corresponding pole in $S_\Delta$. Secondly, the scattering matrix has extra poles at half-integer values of $\Delta$, even in pure hyperbolic space, related to the logarithms that appear in the boundary expansion \ref{boundaryExpansion} when the asymptotic powers differ by an integer, requiring additional counterterms.} In the same way that the resolvent is related to the bulk Green's function, the scattering matrix is related to the CFT two-point function $\langle\op(x_1)\op(x_2)\rangle_\bdry$ in the relevant background. This will also diverge as a function of $\Delta$ as the critical dimension is approached (the familiar divergence in susceptibility at a second-order phase transition), with a pole at $\Delta_c$, signalling the breakdown of the linearized bulk theory when the scalar becomes unstable.

So far we have been quite general.  We will now focus on the case of three dimensions, where we can
make more concrete statements about $\Delta_c$.

\subsection{Locally hyperbolic spaces}

Let us now consider the case where the bulk geometry $\M$ is a locally hyperbolic space of the form 
$\bulk=\quotient{\HH^3}$. We will be primarily interested in the case where $\M$ is handlebody, so we will take $\Gamma$ to be a Schottky group of genus $g>1$. In fact, the results of this section will apply in greater generality, to non-handlebodies, to some geometries containing conical defects, as well as to hyperbolic manifolds of general dimension.\footnote{
The technical assumptions required are only that $\Gamma$ is not elementary, which excludes a few simple cases, most notably the cyclic groups corresponding to the Euclidean BTZ geometry, and that it is geometrically finite, which is true in physically relevant cases and in particular for the Schottky groups with genus $g>1$.}

We will consider a bulk scalar propagating on this geometry, and characterize the relevant spectral theory in terms of properties of the quotient group $\Gamma$.  We will only motivate and explain the results here, referring to the appropriate mathematics literature for more details, precise statements, and proofs.

Consider first computation of the bulk two-point function of $\phi$ in the geometry $\bulk=\quotient{\HH^3}$. 
This can be computed using the method of images, by starting with the two-point function in $\HH^3$ and summing over all elements of the group, corresponding to sources at all image points.  The result is 
\begin{equation}
        G_\Delta^\bulk(y,y') = \sum_{\gamma\in\Gamma} G_\Delta^{\HH^3}(y,\gamma\cdot y') =-\frac{1}{2\pi} \sum_{\gamma\in\Gamma} \frac{e^{-\Delta d(y,\gamma\cdot y')}}{1-e^{-2d(y,\gamma\cdot y')}}
\end{equation}
where $d(y,y')$ is the geodesic distance between the points $y$ and $y'$, with respect to the $\HH^3$ metric. Formally, this gives a function invariant under the group  $\Gamma$, hence well defined on $\bulk$, and solves the Klein-Gordon equation with the appropriate source.  However, this function will not be well-defined if the sum over images fails to converge. In particular, if the number of image points with $d(y,\gamma\cdot y')$ less than some distance $d$ grows rapidly enough as $d\to\infty$ (for some fixed $y,y'$), then the sum will diverge.  More specifically, if the number of image points with $d(y,\gamma\cdot y')<d$ grows like $e^{\delta d}$, then the sum will diverge for $\Delta<\delta$. In fact, this is always the case for some $\delta>0$, as stated in the following result of Sullivan \cite{sullivan1979density}:
\begin{theorem}
        The series
        \begin{equation}
                G_\Delta^\bulk(y,y')= -\frac{1}{2\pi}\sum_{\gamma\in\Gamma} \frac{e^{-\Delta d(y,\gamma\cdot y')}}{1-e^{-2d(y,\gamma\cdot y')}}
        \end{equation}
        converges in the right half-plane $\Re\Delta>\delta$, where $\delta>0$, the \emph{exponent of convergence} of $\Gamma$, is the location of the first resonance of $\quotient{\HH^3}$. The Green's function $G_\Delta^\bulk(y,y')$ (analytically continued in $\Delta$) has a pole at $\Delta=\delta$, and the residue of that pole is given by
        \begin{equation}
                \res_{\Delta\to\delta}G_\Delta^\bulk(y,y') \propto \phi_0(y)\phi_0(y'),
        \end{equation}
        where $\phi_0$ is the zero mode function, the solution of the free bulk wave equation with source-free boundary conditions.
\end{theorem}
As described in the previous section, this pole in the Green's function, the resonance, signals the onset of an instability.  Thus the critical dimension $\Delta_c$ equals the exponent of convergence $\delta$. We emphasize that $\delta$ is strictly positive given our assumptions on $\bulk$, which implies that any handlebody of genus greater than one will be unstable if there is a sufficiently light operator in the spectrum.

We may also compute the CFT two-point function of $\op$ in this background, by taking the limit of the bulk Green's function as the points approach the boundary. The exact result for the two point function will depend on the conformal frame, which corresponds to a choice of regulator as we take the points to the boundary, but the convergence properties of the sum over images will be insensitive to this choice. We can write a general metric on the boundary as $ds^2 = e^{2\sigma(w)}dwd\bar{w}$, where $w$ is the complex coordinate on which $\Gamma$ acts by M\"obius maps. The conformal factor $\sigma$ is defined on the regular set $\Omega$ of $\Gamma$, and defines a metric on the quotient manifold $\bdry = \quotient{\Omega}$ under the condition $\sigma(\gamma(w)) = \sigma(w)-\log|\gamma'(w)|$ for all M\"obius maps $\gamma\in\Gamma$. The bulk computation of the two-point function on $\bdry$ gives a sum over images:
\begin{equation}\label{2ptSum}
        \langle\op(w)\op(w')\rangle_\bdry = e^{-\Delta\sigma(w)} e^{-\Delta\sigma(w')} \sum_{\gamma\in\Gamma}\frac{|\gamma'(w)|^\Delta}{|\gamma(w)-w'|^{2\Delta}}.
\end{equation}
In the summation on the right hand side, the denominator corresponds to the two-point function on the plane, and the numerator is the conformal factor appropriate for each image. 
Once again, this sum converges in the right half-plane $\Re\Delta>\delta$, and the divergence in the two-point function signals the onset of a second-order phase transition. In the mathematical literature, this CFT two-point function appears as the kernel of the scattering matrix \cite{patterson1989selberg}.

This sum can be used to gain some intuition about the relationship between the exponent of convergence $\delta$ and the geometry of the group $\Gamma$.
 The first thing to note is that the tail of the sum, which controls the divergence, is closely related to the limit set $\Lambda$ of the group $\Gamma$.  The limit set is the set of points where the images $\gamma(w)$ accumulate, for any starting point $w$. More precisely, a point is in $\Lambda$ if every neighbourhood of that point contains infinitely many of the images $\gamma(w)$. These are the places where the quotient by $\Gamma$ acts `badly', which we must remove to form the regular set $\Omega=\CC^*-\Lambda$, so we obtain the nice quotient space $\bdry = \quotient{\Omega}$. The tail of the sum is controlled by the limit set, since only a finite number of terms in the sum will lie outside of any arbitrarily small neighbourhood of $\Lambda$. In the simple case of BTZ, $\Gamma$ is the cyclic group consisting of the maps $\gamma_n(w)=q^{2n} w$ for $n\in\ZZ$, and $\Lambda$ consists of the two points $0$ and $\infty$.  In most other cases, however, $\Lambda$ is much more complicated.
 
 For any limit point (that is, element of the limit set), there is a sequence of images of our starting point $w$ that approach it, say $\gamma_n(w)$ for some $\gamma_n\in \Gamma$ (which are independent of $w$). As $n$ increases, $\gamma_n$ will usually be a longer and longer word built out of the generators. For the images to tend to some limit, the $\gamma_n$ must eventually start with the same string of generators, because if they don't, they would map $w$ to places that are separated by some finite distance: as the sequence $\gamma_n$ goes on, the words built out of the generators get longer and longer, and only change later and later on in the string. More precisely, the $k$th letter of the word $\gamma_n$ is constant after some sufficiently large $n$, for any $k$. For each limit point, we can in this way construct a \emph{unique} semi-infinite word built from the generators of the group, a sort of decimal expansion, but using M\"obius maps instead of digits. Such words are in one-to-one correspondence with elements of $\Lambda$. For $g\geq 2$, this set is not only infinite, but uncountable. The `rational numbers' in the analogy with decimals consist of strings of generators that eventually repeat, and are in one-to-one correspondence with the primitive elements $\primes$ of the group $\Gamma$, that is, elements that cannot be written as $\gamma^n$ for any $n>1$ (excepting the identity), corresponding to the attractive fixed point of that element.
 
\begin{figure}
        \centering
        \begin{subfigure}[b]{.45\textwidth}
                \includegraphics[width=.8\textwidth]{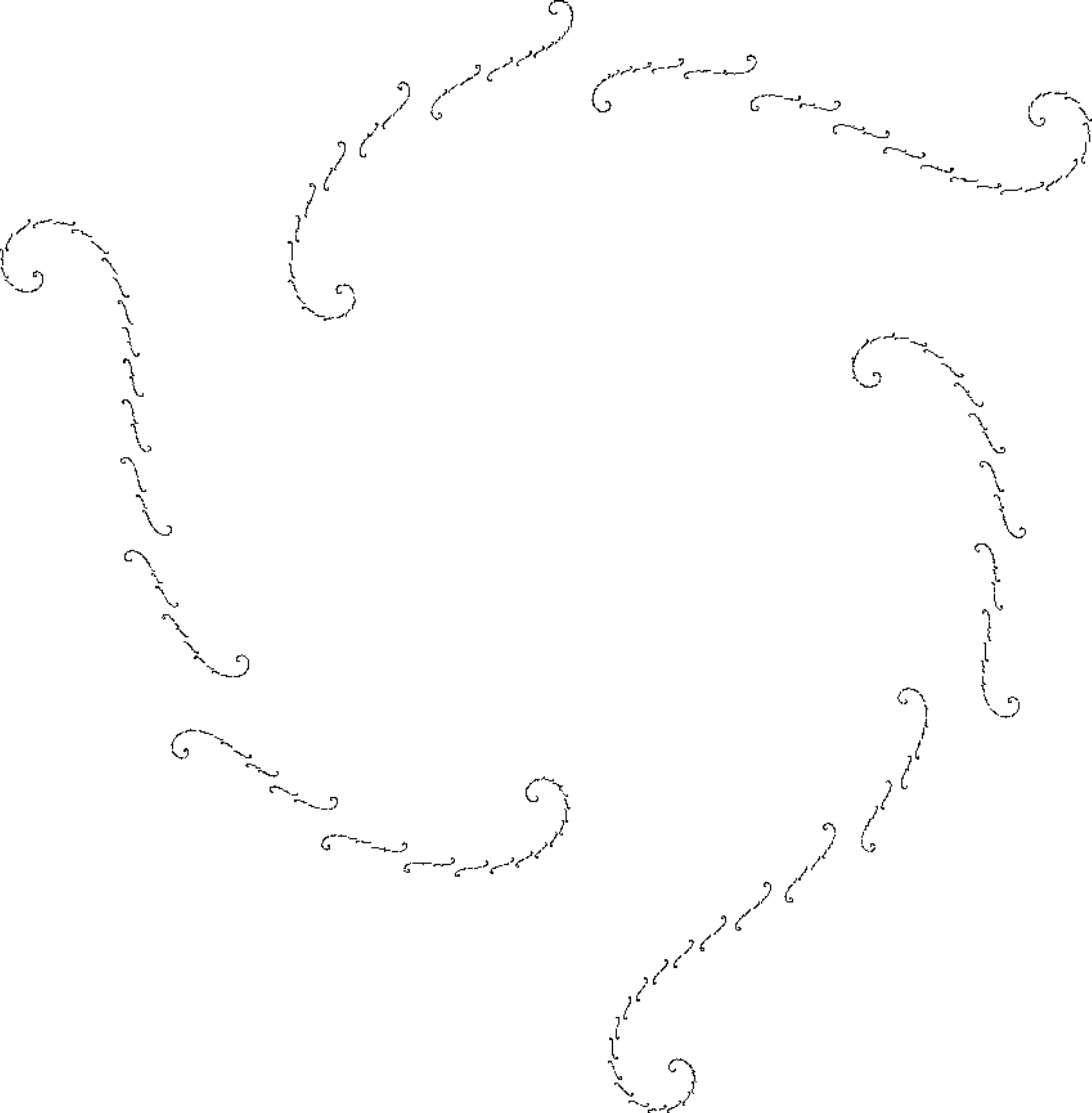}
                        \caption{$q = .8 + .3i$, $\delta\approx 1.09$}
        \end{subfigure}
        \begin{subfigure}[b]{.45\textwidth}
                \includegraphics[width=.8\textwidth]{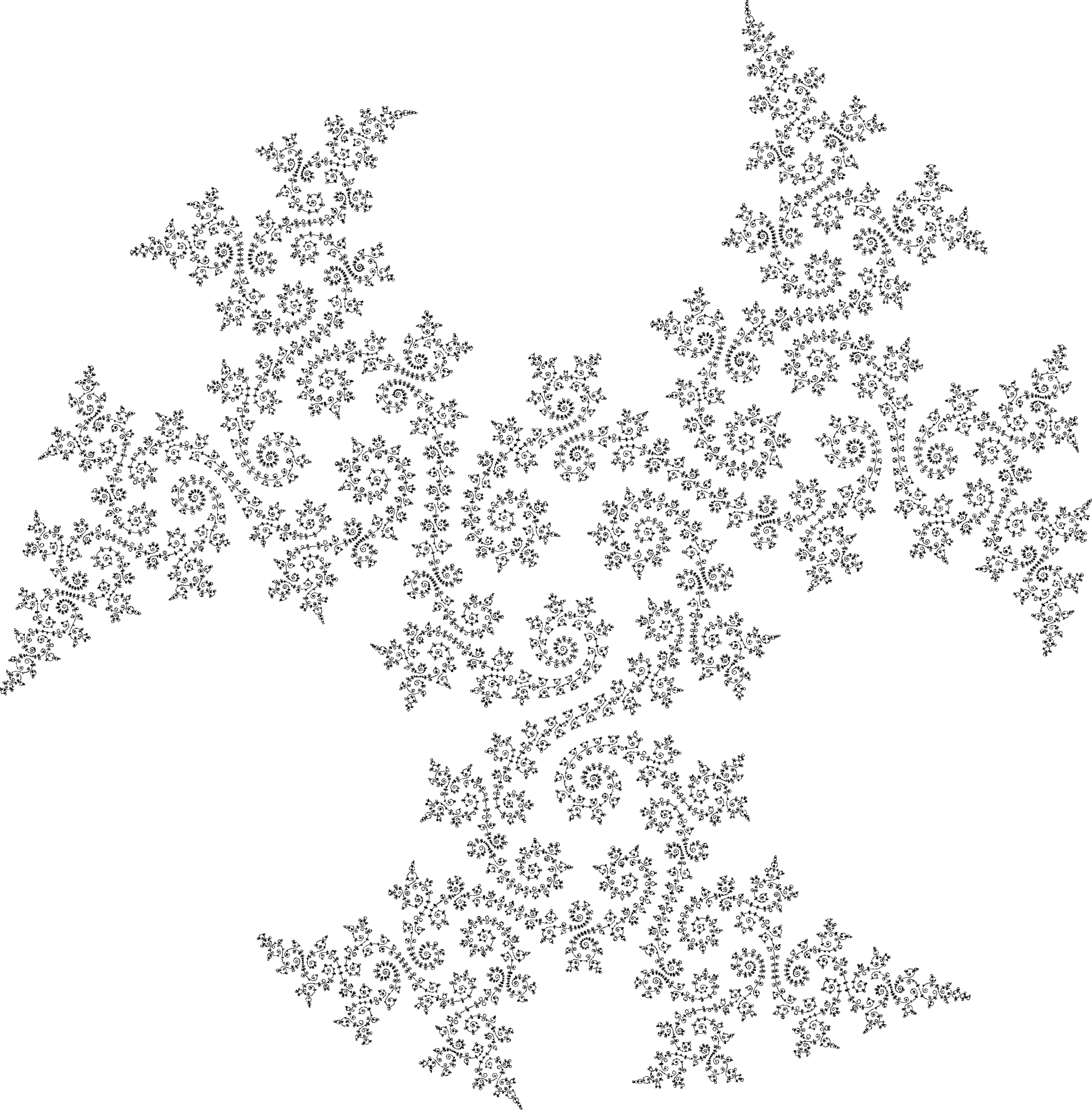}
        \caption{$q = .8 + .44i$, $\delta \approx 1.29$}
        \end{subfigure}
        \caption{ The limit sets for two of the $\ZZ_3$ symmetric genus two Schottky groups that arise when investigating $n=3$ R\'enyi entropies. The parameter $q$ defining the groups is an eigenvalue of one of the generators as specified in \cref{exampleSurfaces}.  We give the value of the Hausdorff dimension $\delta$ for these two limit sets, computed using the methods of \cref{criticalDimSec}.}\label{Limitsets}
\end{figure}

The resulting set $\Lambda$, which controls the tail of the sum over the group, has a rich and beautiful fractal structure. For Fuchsian groups, generated by matrices with real entries, the limit set is a subset of the real line, and closely resembles (indeed, topologically, is homeomorphic to) a Cantor set. Allowing more general Schottky groups, the limit set moves into the complex plane, forming a twisting, intricate, self-similar pattern. Several examples arising when investigating the 3rd R\'enyi entropy of two intervals are illustrated in \cref{Limitsets}. For many more images of limit sets, and a playful semi-popular account of the mathematics involved, we encourage a foray into \cite{mumford2002indra}.

Secondly, we note that the size of the terms in the sum \eqref{2ptSum} is controlled primarily by the factor $|\gamma'(w)|^\Delta$, which describes how things  scale under the action of $\gamma$ (in the flat or round metric on the Riemann sphere, not the metric pulled back from $\bdry$). Given some small set near $w$, the characteristic length of its image under $\gamma$ is scaled by $|\gamma'(w)|$, and its area is scaled by $|\gamma'(w)|^2$, so it is natural generalise this, and say that $|\gamma'(w)|^\Delta$ characterises the scaling in a $\Delta$-dimensional notion of measure, where $\Delta$ can be any positive real number. The convergence of the sum is therefore determined by the trade-off between the accumulation of many points at the limit set, and the shrinking of $\Delta$-dimensional measure associated to images at those points. The critical dimension will occur when these two effects precisely balance, which is when the limit set itself can be assigned a $\Delta$-dimensional measure invariant under $\Gamma$. Hopefully this discussion makes plausible the following theorem of Patterson \cite{patterson1976limit}, Sullivan \cite{sullivan1979density,sullivan1987related} and Bishop-Jones \cite{bishop1997hausdorff}, the precise statement of which uses the notion of \emph{Hausdorff dimension}, a non-integer dimension defined for fractals in metric spaces.
\begin{theorem}[Patterson-Sullivan] \label{PSthm}
        The exponent of convergence $\delta$ is equal to the Hausdorff dimension of the limit set $\Lambda$ of $\Gamma$:
        \begin{equation}
                \delta = \hdim(\Lambda)
        \end{equation}
\end{theorem}
This result connecting spectral theory and fractal geometry is certainly beautiful, which would be justification enough to include it in a mathematics paper, but amazingly enough it is also useful. Firstly, it gives us a new tool to intuit how the critical dimension $\Delta_c$ depends on the geometry, particularly in certain limits. But more importantly, it provides a method to accurately and efficiently compute $\delta$ for any group $\Gamma$, which is far better than na\"ively  solving the bulk Laplace equation numerically, or the method of extracting $\delta$ from the asymptotics of the terms in the sums introduced above. We will discuss an algorithm to compute $\delta$ in \cref{McAlgorithm}, and use it to present both numerical and analytic results.

\subsection{Divergence of the partition function}

In this section we will offer one final perspective on the phase transition, to make a direct connection with the CFT argument discussed in section 3.  In that section we summed up the contributions to the partition function from of a generalized free field, using the global limit of higher genus blocks with the spectrum.
From the bulk point of view, this object is precisely the one-loop partition function of the bulk scalar field $\phi$:
\begin{equation}
        Z_\text{GF}=Z_\text{1-loop} = \frac{1}{\sqrt{\det\left(m^2-\nabla^2\right)}}.
\end{equation}
This makes it apparent that the zero mode should again be visible as a zero eigenvalue of the operator $\nabla^2-m^2$ (defined with suitable boundary conditions).  In this way, the calculations of section 3 put a lower bound on $\Delta_c$.

The bulk computation of this one-loop partition proceeds much as the Green's function computation given above.  
In particular, one can compute this one-loop determinant using heat kernel methods and a sum over images \cite{Giombi:2008vd}.
For higher genus surfaces this one-loop determinant can be written as an infinite product
\begin{equation}\label{bilbo}
        Z^{\quotient{\HH^3}}_\text{1-loop}(\Delta) = \frac{1}{\sqrt{\zeta_\Gamma(\Delta)}},\quad\text{where } \zeta_\Gamma(\Delta) \equiv \prod_{\gamma\in\primes} \prod_{n,\bar{n}=0}^\infty \left(1-q_\gamma^{\Delta+2n} \bar{q}_\gamma^{\Delta+2\bar{n}}\right) \; .
\end{equation}
Here, $q_\gamma$ is the smaller eigenvalue of $\gamma$, as previously introduced. The product is over primitive conjugacy classes $\gamma\in\primes$ of the group $\Gamma$; these are conjugacy classes of elements which cannot be written as a power $\gamma^n$ of another element with $n>1$, and $q_\gamma$ is the smaller of the eigenvalues of $\gamma$ when written as an $SL(2,\CC)$ matrix.  Note that this definition counts $\gamma^{-1}$ separately from $\gamma$, so that terms come in matching pairs. 
We have written the one-loop partition function terms of the \emph{Selberg zeta function} $\zeta_\Gamma$ associated to the group $\Gamma$, as defined in \cite{patterson1989selberg}\footnote{There are several closely related definitions of the Selberg zeta function. The definition we have given is appropriate for hyperbolic three-manifolds where $\Gamma$ is a Kleinian group; another, more common definition is in the context of hyperbolic surfaces, where $\Gamma$ is a Fuchsian group, so $q_\gamma$ is real, and the product over $\bar{n}$ is absent.}. We note that the product in \eqref{bilbo} converges in the same right half-plane $\Re\Delta>\delta$ as the image sums we have already introduced.  In fact, the Selberg zeta function can be analytically continued to an entire function, with zeros precisely at the eigenvalues and resonances of the Laplacian on $\bulk$, as expected \cite{patterson2001divisor}. In particular, the first resonance, corresponding to the phase transition of interest, leads to the one-loop partition function diverging as $(\Delta-\Delta_c)^{-1/2}$. This is the divergence found from the CFT analyses of \cite{Belin:2017nze} and \cref{CFTsec}.

The product \eqref{bilbo} has a simple geometric interpretation in terms of the closed geodesics on the bulk manifold $\bulk=\quotient{\HH^3}$.  Since $\Gamma$ is the fundamental group of $\bulk$, its conjugacy classes are in one-to one correspondence with homotopy classes of closed loops in the bulk, and in a hyperbolic manifold, there is a unique closed geodesic in each class. The primitive conjugacy classes $\primes$ correspond to \emph{prime} geodesics that do not trace over their image multiple times. The geometric parameters associated to a closed geodesic are its length $\ell_\gamma$, and its twist $\theta_\gamma$, the angle through which a normal vector gets rotated after being parallely 
transported around the curve, and are related to the associated conjugacy class of $\Gamma$ by $q_\gamma^2=e^{-\ell_\gamma+i\theta_\gamma}$. The convergence of the product is therefore controlled by the asymptotics of the length spectrum of the bulk manifold. A precise statement of this is given by the prime geodesic theorem, so called because of its close analogy with the prime number theorem (provable using the analytic properties of the Selberg and Riemann zeta functions respectively):
\begin{theorem}[Prime geodesic theorem]
        The prime geodesic counting function $\pi_\bulk(\ell)$, defined as the number of prime geodesics of length at most $\ell$, satisfies the asymptotic formula
        \begin{equation}
                \pi_\bulk(\ell)\sim \frac{e^{\delta\ell}}{\delta\ell} \text{ as }\ell\to\infty.
        \end{equation}
\end{theorem}
In this way, the instability of the scalar field is controlled by the asymptotic properties of the spectrum of very long geodesics. 
This relation between the spectrum of the Laplacian and the lengths of closed geodesics is a special case of the Selberg trace formula (or an appropriate generalization).

Consideration of the partition function leads to an alternative approach to the computation of $\delta$, which we will not pursue further here, by numerically computing the Selberg zeta function, which can be done efficiently (though not directly from the product definition), and locating its zeros.

\subsection{When $\Gamma$ is Fuchsian}\label{Fuchsian}

In the case when the group $\Gamma$ is Fuchsian, i.e.\ when all elements are in $SL(2,\RR)$ and so fix the real line (perhaps after conjugation with some M\"obius map, for example fixing the unit circle instead), the discussion simplifies somewhat. Instead of requiring the full three dimensional geometry, all the main results discussed here can be reduced to the two-dimensional slice $\slice$ fixed by complex conjugation. In this section we briefly describe this reduction and its consequences.

Fuchsian groups are, in many circumstances, the most physically interesting cases, primarily because they correspond to geometries that have a real Lorentzian description. Interpreting the action of complex conjugation as a time-reversal symmetry, the slice $\slice$ fixed by time-reversal has vanishing extrinsic curvature, and hence can be interpreted as an initial Cauchy surface for Lorentzian evolution. Very explicitly, the Euclidean bulk $\bulk$ can be written as
\begin{equation}
        ds^2 = d\chi^2 + \cosh^2 \chi\; d\slice^2 
\end{equation}
where $d\slice^2$ is the hyperbolic metric on the $\chi=0$ slice $\slice$. The Lorentzian geometry (or, rather, a patch of it) is obtained by analytic continuation $\chi\to i t$.  This gives a locally AdS$_3$ solution to the equations of motion in an FRW-like coordinate system, where the spatial slices have constant negative curvature.

It is important to note that while all Fuchsian groups have a reflection symmetry, and corresponding Lorentzian interpretation, the converse is not true: a non-Fuchsian Schottky group may have a time-reflection symmetry and good Lorentzian continuation. To take one example, the pure entangled state on three copies of the CFT obtained by the path integral on a pair of pants is, for certain moduli, dual in the Lorentzian section to disconnected copies of pure AdS and BTZ \cite{Maxfield:2016mwh}, but the corresponding (connected) Euclidean geometry is not described by a Fuchsian group.

The bulk metric is not static, so to simplify the Laplacian it is not as straightforward as choosing a time-independent ansatz. But it is not much harder than that; instead, look for a separable eigenfunction $F(\sigma,\chi)=f(\sigma)g(\chi)$, finding that if $f$ is an eigenfunction of the Laplacian on the $\chi=0$ slice with eigenvalue $\Delta(\Delta-1)$, and obeys the correct AdS boundary conditions, then
\begin{equation}
        F(\sigma,\chi) = (\operatorname{sech}\chi)^{\Delta}\, f(\sigma)
\end{equation}
is  an eigenfunction of the full handlebody Laplacian with eigenvalue $\Delta(\Delta-2)$, with the correct boundary conditions. From this, the critical dimension of the handlebody is determined by the bottom of the spectrum of the slice $\slice$, and computing the actual profile of the zero mode is no longer a three-dimensional problem. 

\section{Results for the critical dimension $\Delta_c$}\label{criticalDimSec}

We have seen that a scalar field on a handlebody $\HH_3/\Gamma$ will be unstable if the dimension is sufficiently small: $\Delta<\Delta_c$. We now turn to an explicit computation of the critical dimension $\Delta_c$, which will be a function of the moduli.   A direct approach, where one studies the Laplacian directly on the geometry of interest, is a complicated numerical task.  Our approach will be to instead use  \cref{PSthm} to calculate $\Delta_c$ from the Hausdorff dimension of the limit set of $\Gamma$. 

Using this, we will obtain analytic results for $\Delta_c$ near the boundary of moduli space, as well as analytic bounds on $\Delta_c$ in the interior of moduli space.  We will also obtain accurate numerical results.  Our main tool will be an algorithm due to McMullen \cite{mcmullen3}, which we now describe.

\subsection{McMullen's algorithm\label{McAlgorithm}}

This section is somewhat technical and is not necessary to understand the results described in later sections.  Readers who are not interested in the details of how the results are obtained can safely skip to section \ref{results}.

To begin, will we need to introduce an additional structure on the limit set $\Lambda$: a $\Gamma$-invariant $\delta$-dimensional measure $\mu$. A measure $\mu$ (though we will not give a precise definition here) allows us to integrate functions on the limit set, in particular assigning a number $\mu(E)=\int_E d\mu\geq 0$ to subsets $E\subseteq\Gamma$, providing a measure of the `content' of $E$. We require the additional property that it transforms as a $\delta$-dimensional density under element of the group\footnote{Here, $|\gamma'|$ computes the local scaling of lengths under the map $\gamma$; we may use any metric on the boundary Riemann sphere for this purpose, for example a round metric, and the result for $\delta$ is insensitive to this choice. For practical computations the flat metric is often most convenient, in which case $|\gamma'|$ is the absolute value of the derivative of the M\"obius map. It is then simplest to require that the point at infinity is not in $\Lambda$, which we will implicitly assume.}:
\begin{equation}\label{gandalf}
 \mu(\gamma(E)) = \int_E |\gamma'|^\delta d\mu \quad\text{for }\gamma\in\Gamma
\end{equation}
A nontrivial measure with this property exists when, and only when $\delta$ equals the Hausdorff dimension of $\Lambda$ (in which case it is unique, up to normalization, for Schottky groups). The only feature of the right hand side that we require is that it is bounded by the measure of the set $\mu(E)$, times the extrema of the integrand $|\gamma'|^\delta$:
\begin{equation}\label{measureBounds}
	\mu(E)\inf_{w\in E}|\gamma'(w)|^\delta \leq \mu(\gamma(E)) \leq \mu(E)\sup_{w\in E}|\gamma'(w)|^\delta
\end{equation}

\begin{figure}[htbp]
        \centering
        \includegraphics[trim = 0 50 0 50, clip, width=.95\textwidth]{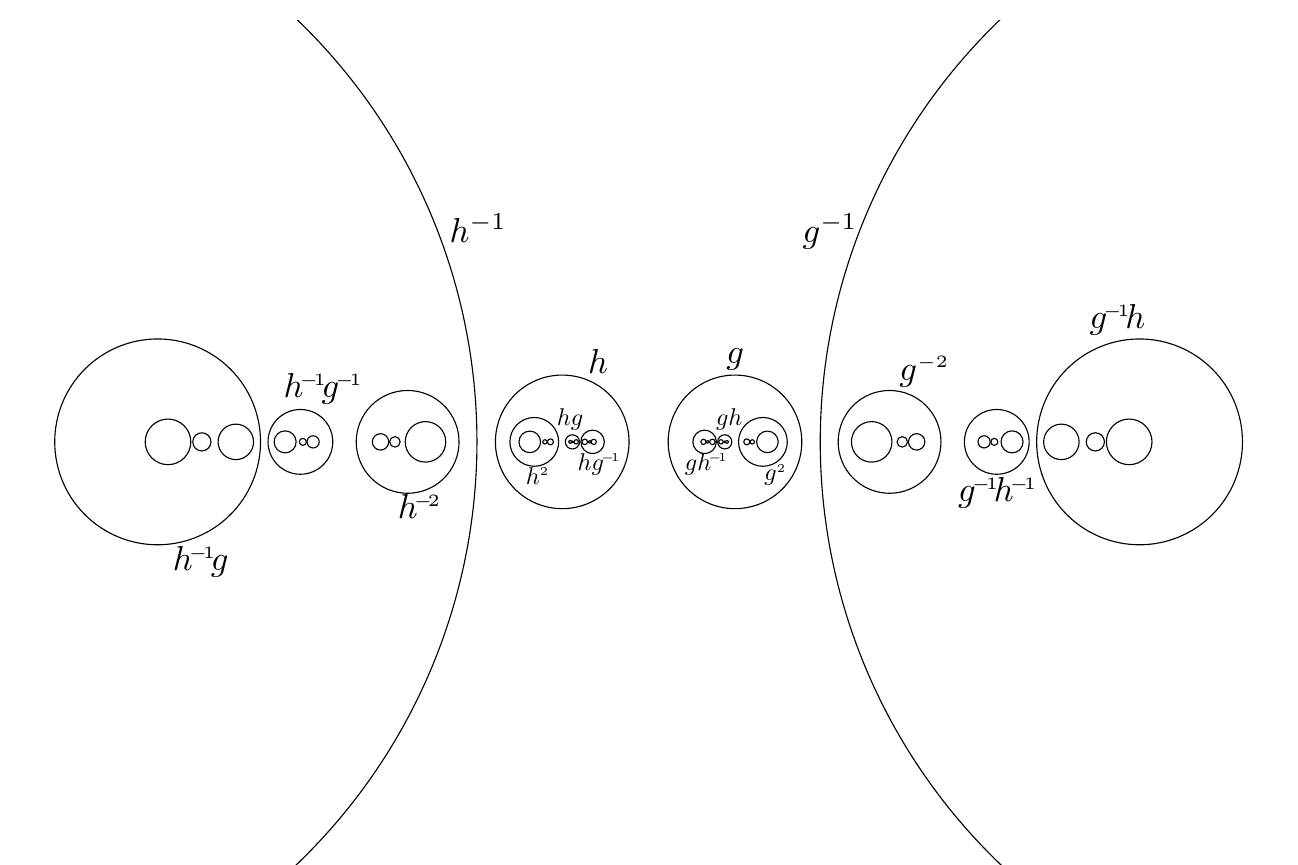}
        \captionsetup{singlelinecheck=off}
        \caption[Example of McMullen's algorithm]{\label{McExample} An example of computing the transition matrix for McMullen's algorithm, in the case of a Kleinian group freely generated by two loxodromic elements $g,h$, so that $\CC^*/\Gamma$ is a genus two surface. In the figure, we have drawn a fundamental domain for $\Gamma$, the exterior of the four outermost circles (those corresponding to $g^{-1}$, $h^{-1}$ are not shown in their entirety). Break the limit set into the four pieces $E_\gamma$ contained in each of these circles, labelled by $\gamma=g,h,g^{-1},h^{-1}$ corresponding to the element of the group that maps the fundamental domain to the interior of the circle, and choose points $w_\gamma\in E_\gamma$, for example the attractive fixed point of $\gamma$. The piece of the limit set $E_g$ can be broken up into three disjoint pieces, inside the circles labelled $g^2$, $gh$ and $gh^{-1}$, which are the images under $g$ of $E_g$, $E_h$ and $E_{h^{-1}}$ respectively. The scalings of these limit sets under the action of $g$ go into the top row of the transition matrix:
        \leavevmode\\\begin{minipage}{\linewidth}
        \vspace{8pt}
                \begin{equation*}
                T = \begin{pmatrix}
                        |g'(w_g)| & |g'(w_h)| & 0 & |g'(w_{h^{-1}})| \\
                        |h'(w_g)| & |h'(w_h)| & |h'(w_{g^{-1}})| & 0 \\
                        0 & |(g^{-1})'(w_h)| & |(g^{-1})'(w_{g^{-1}})| & |(g^{-1})'(w_{h^{-1}})| \\
                        |(h^{-1})'(w_g)| & 0 & |(h^{-1})'(w_{g^{-1}})| & |(h^{-1})'(w_{h^{-1}})|
                \end{pmatrix}
                \end{equation*}
        \vspace{4pt}
        \end{minipage}
        The other three rows repeat the same exercise for the other three regions, and finding $\delta$ such that the spectral radius of $T^\delta$ is unity gives an approximation for the Hausdorff dimension.
        This can be refined by breaking the limit set up into the $3 \times 4^{n-1}$ regions $E_\gamma$ labelled by words of length $n$ in $g,h,g^{-1},h^{-1}$, and applying the $\delta$-invariance imposed by considering the preimage of $E_\gamma$ under the first element ($g,h,g^{-1}$, or $h^{-1}$) appearing in the word $\gamma$. Then $T$ will be a sparse matrix, with three nonzero elements in each row and column, and the algorithm has error decreasing exponentially with $n$. The figure includes labels for words of length two, but also shows the images of circles under words of length three (unlabelled).
        }
\end{figure}

McMullen's algorithm works by splitting the limit set into a finite number of disjoint pieces $E_i$, and attempting to approximate (or bound) the value of the measure $\mu$ on each of these pieces, $\mu_i=\mu(E_i)$. For a detailed explicitly worked example of this and the following, see \cref{McExample}. We begin by imposing \eqref{gandalf}. Specifically, suppose we have some M\"obius map $\gamma_i\in\Gamma$, and one of the pieces $E_i$, whose preimage under $\gamma_i$ is the union of some pieces $E_{j_1},E_{j_2},\dots E_{j_n}$ (one of which may be $E_i$ itself):
\begin{equation}
        E_i = \bigcup_{k=1}^n \gamma(E_{j_k})
\end{equation}
If we pick some points $z_j\in E_j$, then \eqref{gandalf} implies that 
\begin{equation}\label{scalingRelation}
        \mu_i \approx \sum_{k=1}^n |\gamma_i'(w_{j_k})|^\delta \mu_{j_k}\, .
\end{equation}
This is not an exact equality because the scale factor $|\gamma'_i|$ is not constant on the limit set.  However, by taking the $E_i$ to be small $|\gamma'_i|$ will be approximately constant, so the error will be small. To be more precise, we may replace the factors of $|\gamma_i'(w_j)|$ by upper or lower bounds on this scaling over the set $E_j$, and use \cref{measureBounds} to replace the approximate equation by inequalities.

With an appropriately chosen partition $\{E_i\}$ and maps $\gamma_i$, a similar argument can be repeated for every $i$.  Our approximate formula \eqref{scalingRelation} can then be written in terms of a square matrix $T$, the \emph{transition matrix}, whose entries $T_{ij}$ are equal to $|\gamma_i'(z_j)|$ for each of the $E_j$ in the preimage $\gamma_i^{-1}(E_i)$, and zero otherwise. Invariance of the measure is then the statement that $\mu_i$ is a unit eigenvector of $T^\delta$, where the power is taken element-wise. Since the matrix $T^\delta$ has 
nonnegative entries it is guaranteed to have a unique eigenvector with positive components; furthermore this is the eigenvector with largest eigenvalue.\footnote{This is the \href{http://lmgtfy.com/?q=perron-frobenius+theorem}{Perron-Frobenius theorem}. Since some of the entries of $T$ are zero, we must also require that $T$ is irreducible.  This means, roughly speaking, that  when we apply $T$ repeatedly all of the regions $E_i$ will eventually mix.} If $\mu$ is to satisfy the invariance criterion, this eigenvalue should be one.  Thus to find the Hausdorff dimension, we find the value of $\delta$ such that the largest eigenvalue of $T^\delta$ equals one.\footnote{This $\delta$ exists and is unique since the spectral radius of $T^\delta$ decreases monotonically as $\delta$ increases from zero to infinity.} An analogous result holds when we replace the approximate equations by inequalities, so by choosing upper or lower bounds on the transition matrix elements, we can obtain rigorous bounds on $\delta$.

For a given partition $\{E_i\}$ of the limit set this gives an estimate for $\delta$, and to obtain a more accurate estimate we can refine the partition into a larger number of pieces. With an appropriate refinement, the result converges rapidly to the Hausdorff dimension, and in practice it is sufficient to use a rather coarse partition of $\Lambda$.  Although this discussion is rather abstract, the explicit implementation of this algorithm is quite straightforward; see figure \ref{McExample} for a simple example.

\subsection{Analytic results}\label{results}

Our first analytic results are for Fuchsian groups; this includes the surfaces described in \cref{exampleSurfaces} with real cross-ratio $0<x<1$. 
In this case we note that the limit points must all lie on the real axis of the $w$-plane, corresponding to the slice fixed by time-reflection symmetry. Since $\Lambda$ is a subset of the one-dimensional line, it must have dimension $\delta \leq 1$. Thus $\Delta_c\le 1$.  So the only potentially unstable fields are those with $\Delta <1$, which correspond to bulk scalars which are quantized using alternate boundary conditions.

In the rest of this section we will focus on the case of the genus $g=n-1$ surfaces with $\ZZ_n$ symmetry, described in \cref{exampleSurfaces}, relevant for computing the $n$th R\'enyi entropy of a pair of intervals. We begin by applying the above algorithm in this case at the coarsest level of approximation, to obtain analytic bounds. These are especially useful at the edge of moduli space where $x\to 0$, because in this limit the different pieces of the limit set become well-separated and small, so the scale factor does not vary much over it. These bounds thus become tighter and tighter as $x\to 0$.

As described in \cref{exampleSurfaces}, we can extend the group $\Gamma$ to $\hat{\Gamma}$, generated by $R:w\mapsto e^{2\pi i/n} w$ and $S:w\mapsto \frac{w-\zeta}{\zeta w-1}$, by including a dihedral group of holomorphic automorphisms. This extension of the group does not alter the limit set, and a $\Gamma$-invariant measure constructed on it will also be invariant under $\hat{\Gamma}$. Using this, we will divide the limit set into $n$ pieces, all related by the $\ZZ_n$ symmetry $R$, and hence having equal measure, and use the mapping under $S$ to constrain the dimension of this measure. This is somewhat simpler than using the original presentation of $\Gamma$ such as in \cref{McExample}.

There are $n$ pieces of the limit set $E_k$, each centred at a root of unity $e^{2\pi i k/n}$, with size of order $\theta^2$ for small $\theta$, and related to each other by $R$. A simple way to show this is by constructing a fundamental domain for $\hat{\Gamma}$, bounded by the radial lines from the origin at angles $\pm\pi i/n$, related by action of $R$, and a circle $C_n$ mapped to itself by $S$, centred at $\sec\theta$ with radius $\tan\theta$ (recall $\zeta=\cos\theta$). Then $E_n$ is the part of the limit set inside this circle, and the remaining $E_k$ are inside corresponding circles $C_k=R^k(C_n)$ obtained by rotations by angle $2\pi k/n$. This immediately bounds the size of the limit set by the size of the circles $C_k$, of order $\theta$, which is a sufficiently strong result for our immediate purposes\footnote{To obtain an improved bound, note that $E_n$ must be contained in a smaller set, namely the union of the interiors of $S(C_k)$, which is concentrated in a region with size of order $\theta^2$.}.

 By the $\ZZ_n$ symmetry, and the fact that the action of $R$ doesn't scale ($|R'|=1$), the sets $E_k$ must all have equal measure $\mu_k=\mu$. We can then apply the action of $S$, which maps $E_1,E_2,\cdots, E_{n-1}$ onto $E_n$. The amount by which $E_k$ scales under $S$ can be computed from
\begin{equation}
        S'(e^{2\pi i k/n}+O(\theta^2))=\frac{\theta^2}{(e^{2\pi i k/n}-1)^2}+O(\theta^4)
\end{equation}
so \cref{scalingRelation}, equating the sum of the scaled measures of $E_1,E_2,\cdots, E_{n-1}$ to the measure of $E_n$, gives us
\begin{equation}\label{znscaling}
        \sum_{k=1}^{n-1} \left(\frac{\theta}{2\sin\frac{\pi k}{n}}\right)^{2\delta}= 1+ O(\theta^2).
\end{equation}
This requires $\delta$ to tend to zero as $\theta$ goes to zero, and solving to leading order for small $\delta$, we get the result that
\begin{equation}
        \delta\sim \frac{\log(n-1)}{2\log |\theta|^{-1}} + O\left(\frac{1}{(\log|\theta|)^2}\right) \quad \text{as }\theta\to 0
\end{equation}
with the first corrections coming from solving the equation to higher order in $\delta$, rather than the order $\theta^2$ variation of the scaling relation giving the correction to \cref{znscaling}. With only minor modifications, this derivation continues to apply if we allow $\theta$ to be complex, which is why we have included the modulus in the result. The case $n=3$, with real $\theta$, was treated in \cite{mcmullen3}, though instead of using $S$, that paper uses a reflection, which requires $\theta$ to be real\footnote{For comparison, the parameter $\theta$ used in that paper is half of the $\theta$ used here.}.

To facilitate comparison between different values of $n$, we write this in terms of the cross-ratio $x$. With the monodromy methods outlined in \cref{exampleSurfaces}, the map from $x$ to the Schottky parameter $\theta$ can be computed as a series expansion, with the leading order result that $\theta = \frac{\sqrt{x}}{n}(1+O(x))$:
\begin{equation}
        \delta= \frac{\log(n-1)}{\log|x|^{-1}}+O\left(\frac{1}{(\log |x|)^2}\right) \quad \text{as }x\to 0
\end{equation}

Note that this is the asymptotic behaviour for fixed $n$ as $x\to 0$, but clearly must break down if $n$ is parametrically large. In the first instance, it is not self-consistent if $n x$ is of order one, since in that case the leading order term in the expansion would be of order one. But we can take a different order of limits to understand what happens at large $n$.

Starting with the scaling relation \cref{znscaling} at small fixed $x$, and na\"ively taking the large $n$ limit term by term, we arrive at
\begin{equation}
        2\sum_{k=1}^{\infty} \left(\frac{\sqrt{x}}{2\pi k}\right)^{2\delta} \approx 1
\end{equation}
where the factor of two is to count contributions both from fixed $k$ and fixed $n-k$. The first thing to notice is that the tail of the series decays as $k^{-2\delta}$, so convergence of the sum immediately requires $\delta>\frac{1}{2}$. Summing the series, we arrive at
\begin{equation}
        \frac{1}{2}\left(\frac{4\pi^2}{x}\right)^\delta \approx \zeta(2\delta)
\end{equation}
from which we see that the left hand side is large for small $x$, since $\delta$ cannot be small. The zeta function must therefore be close to the pole at $\delta=\frac{1}{2}$, and we can find a perturbative solution:
\begin{equation}
        \lim_{n\to\infty}\delta = \frac{1}{2} +\frac{\sqrt{|x|}}{2\pi} + O(x)
\end{equation}
This result should be interpreted as the limit of $\delta$ as $n\to\infty$, for fixed but small $x$. 
It turns out this na\"ive argument is, in essence, correct, and can be made completely precise by repeating the argument for the group generated by one parabolic element and one elliptic element of order 2, equivalent to Theorem 3.6 of \cite{mcmullen3} (for real cross-ratio). In fact, the result that the Hausdorff dimension does not go to zero as $x\to 0$ is a consequence of a general result, that $\delta>\frac{1}{2}$ whenever the group in question contains a parabolic element (Corollary 2.2 in \cite{mcmullen1}).

This Schottky group with $R$ parabolic, instead of elliptic order $n$, corresponds to the `$n=\infty$' version of the geometry described in \cref{exampleSurfaces}. This complex one-dimensional family of Kleinian groups is known in the mathematics literature as the `Riley slice' of Schottky space. It is a little tricky to think about the $n\to\infty$ limit of the handlebody, bounded by a Riemann surface of infinite genus, but it is rather simpler to understand the geometry after taking a quotient by the $\ZZ_n$ replica symmetry, as suggested in \cite{Lewkowycz:2013nqa}. The boundary of this geometry is just the original Riemann sphere, and the bulk has conical defects, of opening angle $2\pi/n$, going from $0$ to $x$ and from $1$ and $\infty$. Taking a formal analytic continuation of the geometry to $n=1$, the conical defects become the Ryu-Takayanagi surface, lying on geodesics \cite{Lewkowycz:2013nqa}, but we are taking the opposite limit, in which the defects become cusps, in particular receding to infinite proper distance.

At this point, let us pause briefly to 
understand the physical consequences. This result means that the $x\to 0$ and $n\to \infty$ limits of the critical dimension do not commute, so that while for any fixed $n$, any dimension of scalar will be stable for sufficiently small $x$, if there is a scalar of dimension less than $1/2$, it will be subject to the phase transition for \emph{any} cross ratio, if $n$ is taken sufficiently large. Note that this is all in a limit where we have taken $c$ to infinity first, and new behaviour dominated by quantum corrections may take over when $n$ is parametrically large in $c$. In particular, the large $n$ limit of the R\'enyi entropy is controlled by the largest eigenvalue of the reduced density matrix, or the `ground state energy' of the modular Hamiltonian $H_A=-\log\rho_A$, but it is unclear whether the semiclassical description is sensitive to a single lowest eigenvalue, or a dense collection of parametrically many low-lying eigenvalues of $H_A$.

To conclude the discussion of analytic results, let us briefly describe the other limit of the geometry, when $x\to 1$, corresponding to the horizon sizes of the multi-boundary wormhole becoming small. This limit is of less direct physical interest, since it is well past the first-order phase transition of the partition function, so is not the dominant saddle-point geometry, but is nonetheless useful to hone intuition. In the case that $x$ is real, so $\Gamma$ is a Fuchsian group, it is sufficient to describe the bulk geometry in terms of the two-dimensional hyperbolic surface making up the $t=0$ slice. In this limit, it is helpful to separate the geometry into the exterior pieces, lying between each boundary and a horizon, and the `convex core' or `causal shadow' region linking them together, bounded by the $n$ horizons. In the $x\to 1$ limit, the horizons become very small, and the centre of the geometry recedes down a long, narrowing neck. The core then approximates the geometry of the negatively curved metric on some compact surface with punctures, though the punctures do not quite pinch off, rather reaching a minimum radius at the narrow horizons where they join to the exterior funnels. On such a surface, the critical dimension approaches one, the maximal eigenvalue of the Laplacian (on the $t=0$ slice, not the Laplacian in the three-dimensional bulk: see \cref{Fuchsian}) being small and positive. The corresponding eigenfunction is roughly constant on the convex core, and small in the exterior funnels, with an interpolation over the long, narrow necks connecting them. The asymptotic behaviour of $\Delta_c$ in this limit can be computed by directly approximating this Laplace eigenfunction (the zero mode of the instability) \cite{dodziuk1987estimating}.

Taking $x\to 1$ through complex values is much more complicated, so we will not be able to say much about it. Schottky space, perhaps parameterized in this case by the values of $\zeta$ corresponding to some complex $x$, itself has a complicated fractal boundary, and the features of the handlebody depend sensitively on how this boundary is approached. This is a deep and beautiful subject, but goes far beyond the scope of this article. In any case, the numerical computations we describe next show that it is possible to obtain dimensions $\delta>1$ in this limit, for example the limit sets illustrated in \cref{Limitsets}.

\subsection{Numerical results}

These analytic results are very useful to understand the behaviour of the critical dimension at the edges of moduli space, and as genus is varied, but McMullen's algorithm is also useful to quickly compute the Hausdorff dimension numerically, to many digits of precision. We conclude by presenting the results of these computations.

\begin{figure}
        \centering
        \includegraphics[width=400pt]{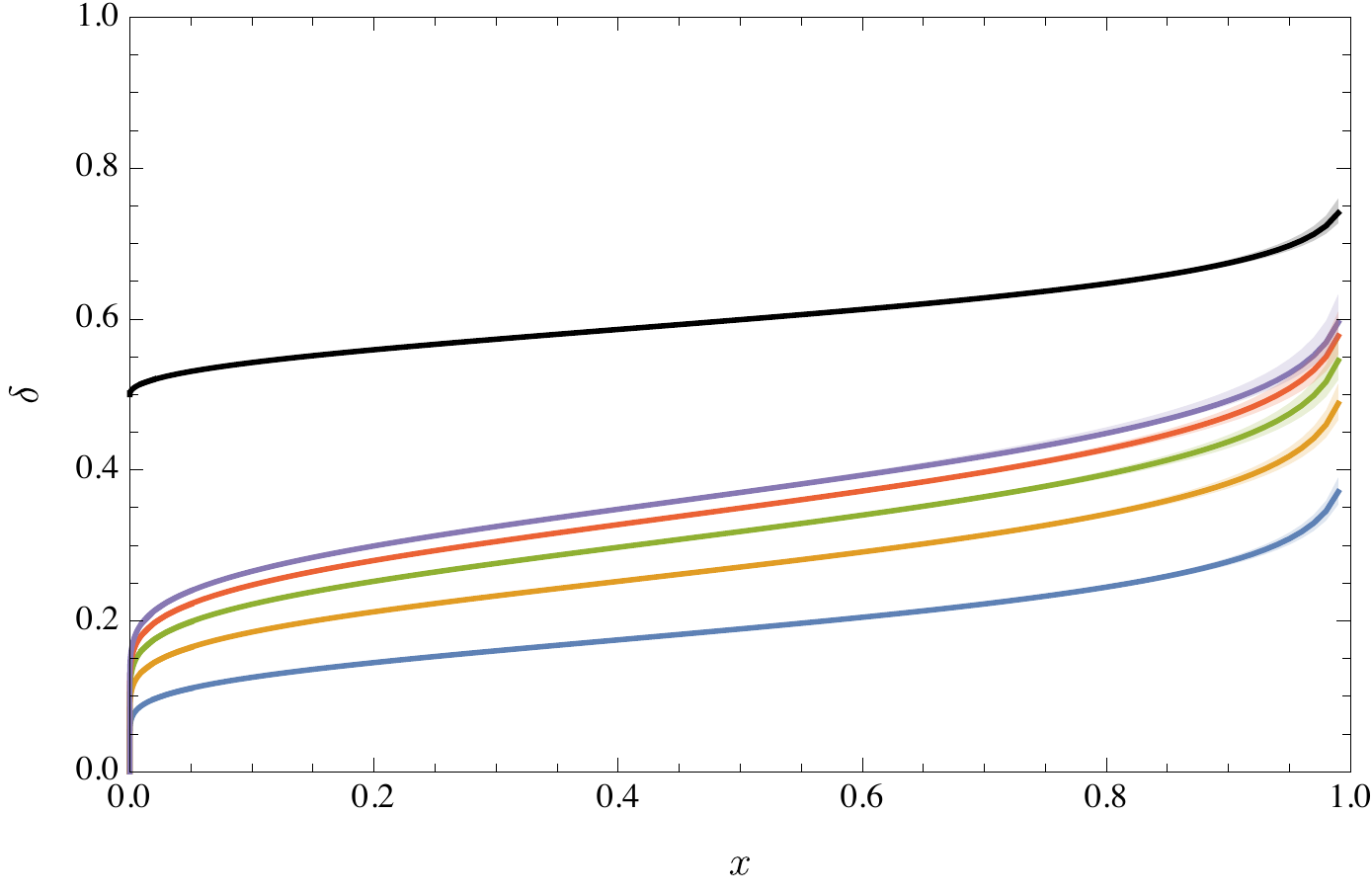}
        \caption{\label{Hausforffx}The critical dimension $\Delta_c=\delta$ as a function of cross-ratio $x$ for the handlebodies corresponding to the R\'enyi entropies of a pair of intervals. From top to bottom, the curves correspond to genus $2,3,4,5,6$, and finally the $n\to\infty$ result in black. The shading visible on the right side of the plot indicates the bounds achieved by applying McMullen's algorithm at the crudest level of approximation.}
\end{figure}

Firstly, \cref{Hausforffx} plots the Hausdorff dimension as a function of cross-ratio $x$ for various values of $n$, including also the limit as $n\to\infty$. With this parameterization, for generic values of $x$, the convergence of the algorithm is remarkably rapid. Indeed, the plot includes shaded regions to indicate the rigorous bounds obtained by applying the algorithm at the crudest level. These are computed by numerically solving the equation \cref{znscaling}, but replacing the terms on the left hand side with upper or lower bounds for $|S'(w)|^\delta$ over $w\in E_k$, rather than the estimates used there. The allowed regions for $\delta$ are in many cases not even visible until $x$ is rather close to 1. Refining further, the algorithm gives results with ten or more digits of precision in a fraction of a second on a laptop of modest specifications. In fact, by far the larger source of computing time and error comes from the conversion between the cross-ratio and Schottky variables, rather than the algorithm to compute $\delta$ from the group generators.

\begin{figure}
        \centering
        \begin{subfigure}[b]{.5\textwidth}
                \includegraphics[width=200pt]{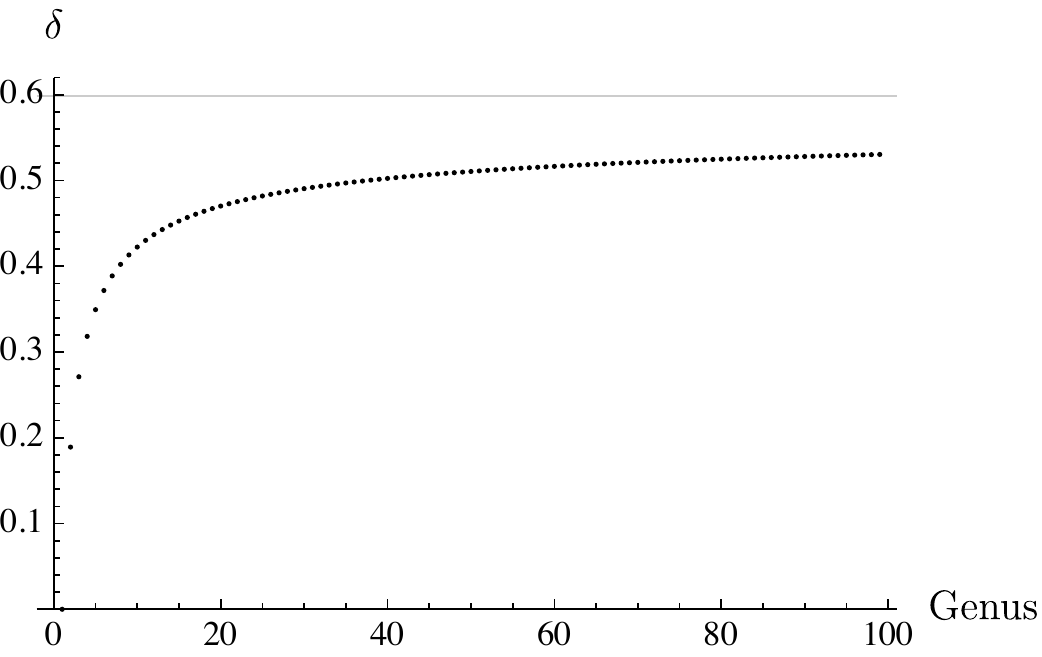}
        \end{subfigure}
        \begin{subfigure}[b]{.45\textwidth}
                \includegraphics[width=200pt]{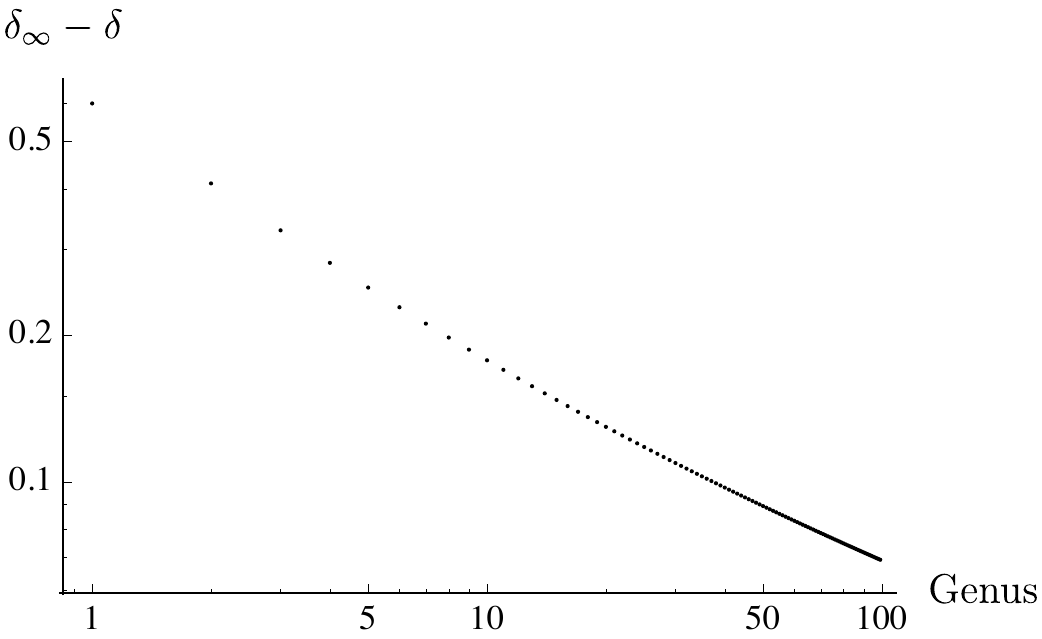}
        \end{subfigure}
        \caption{The critical dimension for the $x=\frac{1}{2}$ R\'enyi surface as a function of replica number $n$. The asymptote is the computed limit as $n\to\infty$. On the right is a log-log plot showing convergence to this value.\label{deltavsn}}
\end{figure}

A physically motivated value to consider is at the boundary with the Hawking-Page phase transition $x=\frac{1}{2}$, which will give the maximum value of $\Delta_c$ within this class of geometries, while in the dominant phase. At genus 2 ($n=3$), 
this value is $\Delta_c=0.189124003$, which is rather close to (and the correct side of) the bound $\Delta_c\geq 0.18912109$ obtained in \cref{CFTsec} from refining the CFT methods of \cite{Belin:2017nze}. Staying at $x=\frac{1}{2}$ and increasing the genus, we find that the critical dimension increases rapidly at first, before slowly approaching the limiting value $\Delta_c\to 0.599$ as $n\to\infty$, as shown in \cref{deltavsn}.

\begin{figure}
        \centering
        \includegraphics[width=.7\textwidth]{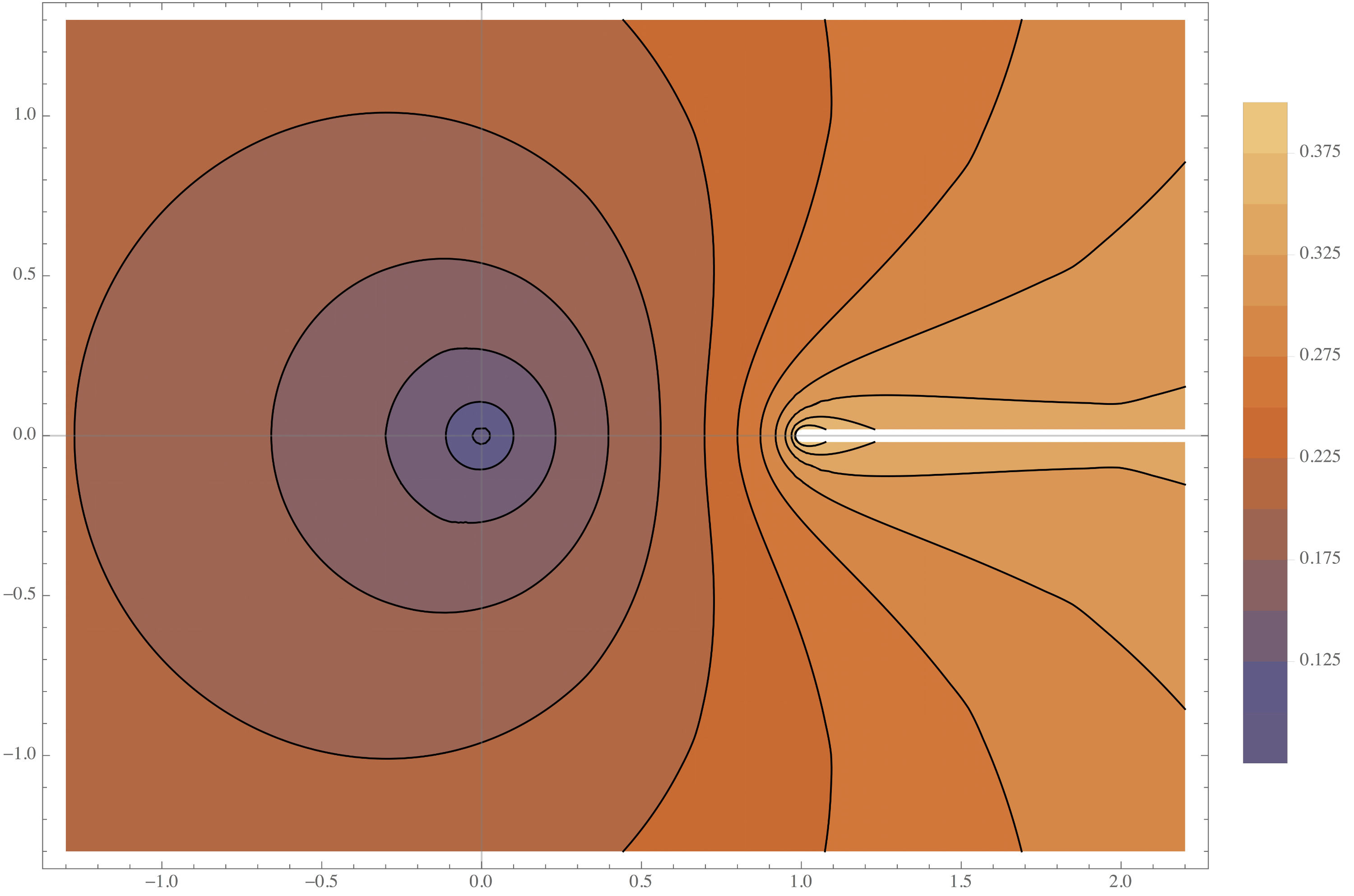} 
        \caption{The Hausdorff dimension of the $\ZZ_3$ symmetric genus two handlebody, as a function of the (complex) cross-ratio $x$.  Note that the Hausdorff dimension goes to zero at the origin ($x=0$) and approaches one as $x\to 1^-$ along the real axis.\label{ContourPlot}}
\end{figure}

In our last plot, \cref{ContourPlot}, we indicate how the Hausdorff dimension behaves for complex values of the cross-ratio, for genus two. Note that this is invariant under inversion in the circle of unit radius centred at one. This is because the extended groups $\hat{\Gamma}$ corresponding to these geometries are the same, though $\Gamma$ consists of different subgroups in each case. From the geometric point of view, taking the $\ZZ_n$ quotient of the handlebody gives the same geometry, with conical defects at the fixed points, though the original geometries are distinct (being branched around the defects in different ways). This relates a cross-ratio $0<x<1$ with a negative cross-ratio $-\frac{x}{1-x}$, which corresponds to swapping the location of twist and anti-twist operators, relevant for computing R\'enyi negativity of two disjoint intervals \cite{Calabrese:2012ew}. The correspondence between the geometries implies a correspondence between the classical limits of R\'enyi entropy and R\'enyi negativity for two intervals, and taking an analytic continuation to $n=1$, the logarithmic negativity of two intervals must vanish (to leading order in $c$) in the regime $x<\frac{1}{2}$ where this geometry dominates the path integral.

Finally, it is interesting to ask what the largest possible value of $\Delta_c$ could be for a geometry that dominates the path integral. A lower bound (precluding surprising new symmetry-breaking phases that dominate the path integral) comes from our numerics for the infinite genus limit, giving examples where $\Delta_c$ as large as $.599$ can be achieved. An interesting result that may bound this in the other direction comes from \cite{hou2016smooth}, showing in particular that every Riemann surface admits a uniformization by a Schottky group of Hausdorff dimension less than one. As a heuristic, matching our expectations in limits of moduli space, the dominant saddle-point seems to be that with minimal Hausdorff dimension, so this is suggestive, though not conclusive, that there may never be a dominant geometry with $\Delta_c>1$.

\acknowledgments

We are very grateful to A. Belin, C. Keller and I. Zadeh for useful conversations.  
We acknowledge the support of the Natural Sciences and Engineering Research Council of Canada (NSERC), funding reference number SAPIN/00032-2015.
This work was supported in part by a grant from the Simons Foundation (385602, A.M.).
This work was performed in part at the Aspen Center for Physics, which is supported by National Science Foundation grant PHY-1607611.

\bibliographystyle{ssg}
\bibliography{biblio}

\begingroup\raggedright\begin{thebibliography}{10}

\bibitem{Maldacena:1997re}
J.~M. Maldacena, ``{The Large N limit of superconformal field theories and
  supergravity},'' {\em Int. J. Theor. Phys.} {\bf 38} (1999) 1113--1133,
  \href{http://xxx.lanl.gov/abs/hep-th/9711200}{{\tt hep-th/9711200}}. [Adv.
  Theor. Math. Phys.2,231(1998)].

\bibitem{Banados:1992wn}
M.~Banados, C.~Teitelboim, and J.~Zanelli, ``{The Black hole in
  three-dimensional space-time},'' {\em Phys. Rev. Lett.} {\bf 69} (1992)
  1849--1851, \href{http://xxx.lanl.gov/abs/hep-th/9204099}{{\tt
  hep-th/9204099}}.

\bibitem{Strominger:1997eq}
A.~Strominger, ``{Black hole entropy from near horizon microstates},'' {\em
  JHEP} {\bf 02} (1998) 009, \href{http://xxx.lanl.gov/abs/hep-th/9712251}{{\tt
  hep-th/9712251}}.

\bibitem{Brown:1986nw}
J.~D. Brown and M.~Henneaux, ``{Central Charges in the Canonical Realization of
  Asymptotic Symmetries: An Example from Three-Dimensional Gravity},'' {\em
  Commun. Math. Phys.} {\bf 104} (1986) 207--226.

\bibitem{Maldacena:1998bw}
J.~M. Maldacena and A.~Strominger, ``{AdS(3) black holes and a stringy
  exclusion principle},'' {\em JHEP} {\bf 12} (1998) 005,
  \href{http://xxx.lanl.gov/abs/hep-th/9804085}{{\tt hep-th/9804085}}.

\bibitem{Dijkgraaf:2000fq}
R.~Dijkgraaf, J.~M. Maldacena, G.~W. Moore, and E.~P. Verlinde, ``{A Black hole
  Farey tail},'' \href{http://xxx.lanl.gov/abs/hep-th/0005003}{{\tt
  hep-th/0005003}}.

\bibitem{Maloney:2007ud}
A.~Maloney and E.~Witten, ``{Quantum Gravity Partition Functions in Three
  Dimensions},'' {\em JHEP} {\bf 02} (2010) 029,
  \href{http://xxx.lanl.gov/abs/0712.0155}{{\tt 0712.0155}}.

\bibitem{Hawking:1982dh}
S.~W. Hawking and D.~N. Page, ``{Thermodynamics of Black Holes in anti-De
  Sitter Space},'' {\em Commun. Math. Phys.} {\bf 87} (1983) 577.

\bibitem{Brill:1995jv}
D.~R. Brill, ``{Multi - black hole geometries in (2+1)-dimensional gravity},''
  {\em Phys. Rev.} {\bf D53} (1996) 4133--4176,
  \href{http://xxx.lanl.gov/abs/gr-qc/9511022}{{\tt gr-qc/9511022}}.

\bibitem{Aminneborg:1997pz}
S.~Aminneborg, I.~Bengtsson, D.~Brill, S.~Holst, and P.~Peldan, ``{Black holes
  and wormholes in (2+1)-dimensions},'' {\em Class. Quant. Grav.} {\bf 15}
  (1998) 627--644, \href{http://xxx.lanl.gov/abs/gr-qc/9707036}{{\tt
  gr-qc/9707036}}.

\bibitem{Krasnov:2000zq}
K.~Krasnov, ``{Holography and Riemann surfaces},'' {\em Adv. Theor. Math.
  Phys.} {\bf 4} (2000) 929--979,
  \href{http://xxx.lanl.gov/abs/hep-th/0005106}{{\tt hep-th/0005106}}.

\bibitem{Headrick:2010zt}
M.~Headrick, ``{Entanglement Renyi entropies in holographic theories},'' {\em
  Phys. Rev.} {\bf D82} (2010) 126010,
  \href{http://xxx.lanl.gov/abs/1006.0047}{{\tt 1006.0047}}.

\bibitem{Cardy:2017qhl}
J.~Cardy, A.~Maloney, and H.~Maxfield, ``{A new handle on three-point
  coefficients: OPE asymptotics from genus two modular invariance},'' {\em
  JHEP} {\bf 10} (2017) 136, \href{http://xxx.lanl.gov/abs/1705.05855}{{\tt
  1705.05855}}.

\bibitem{Cho:2017fzo}
M.~Cho, S.~Collier, and X.~Yin, ``{Genus Two Modular Bootstrap},''
  \href{http://xxx.lanl.gov/abs/1705.05865}{{\tt 1705.05865}}.

\bibitem{Keller:2017iql}
C.~A. Keller, G.~Mathys, and I.~G. Zadeh, ``{Bootstrapping Chiral CFTs at Genus
  Two},'' \href{http://xxx.lanl.gov/abs/1705.05862}{{\tt 1705.05862}}.

\bibitem{Balasubramanian:2014hda}
V.~Balasubramanian, P.~Hayden, A.~Maloney, D.~Marolf, and S.~F. Ross,
  ``{Multiboundary Wormholes and Holographic Entanglement},'' {\em Class.
  Quant. Grav.} {\bf 31} (2014) 185015,
  \href{http://xxx.lanl.gov/abs/1406.2663}{{\tt 1406.2663}}.

\bibitem{Yin:2007gv}
X.~Yin, ``{Partition Functions of Three-Dimensional Pure Gravity},'' {\em
  Commun. Num. Theor. Phys.} {\bf 2} (2008) 285--324,
  \href{http://xxx.lanl.gov/abs/0710.2129}{{\tt 0710.2129}}.

\bibitem{Yin:2007at}
X.~Yin, ``{On Non-handlebody Instantons in 3D Gravity},'' {\em JHEP} {\bf 09}
  (2008) 120, \href{http://xxx.lanl.gov/abs/0711.2803}{{\tt 0711.2803}}.

\bibitem{Maxfield:2016mwh}
H.~Maxfield, S.~Ross, and B.~Way, ``{Holographic partition functions and phases
  for higher genus Riemann surfaces},'' {\em Class. Quant. Grav.} {\bf 33}
  (2016), no.~12 125018, \href{http://xxx.lanl.gov/abs/1601.00980}{{\tt
  1601.00980}}.

\bibitem{Gubser:2008px}
S.~S. Gubser, ``{Breaking an Abelian gauge symmetry near a black hole
  horizon},'' {\em Phys. Rev.} {\bf D78} (2008) 065034,
  \href{http://xxx.lanl.gov/abs/0801.2977}{{\tt 0801.2977}}.

\bibitem{Hartnoll:2008vx}
S.~A. Hartnoll, C.~P. Herzog, and G.~T. Horowitz, ``{Building a Holographic
  Superconductor},'' {\em Phys. Rev. Lett.} {\bf 101} (2008) 031601,
  \href{http://xxx.lanl.gov/abs/0803.3295}{{\tt 0803.3295}}.

\bibitem{Belin:2014lea}
A.~Belin and A.~Maloney, ``{A New Instability of the Topological black hole},''
  {\em Class. Quant. Grav.} {\bf 33} (2016), no.~21 215003,
  \href{http://xxx.lanl.gov/abs/1412.0280}{{\tt 1412.0280}}.

\bibitem{Calabrese:2004eu}
P.~Calabrese and J.~L. Cardy, ``{Entanglement entropy and quantum field
  theory},'' {\em J. Stat. Mech.} {\bf 0406} (2004) P06002,
  \href{http://xxx.lanl.gov/abs/hep-th/0405152}{{\tt hep-th/0405152}}.

\bibitem{Belin:2013dva}
A.~Belin, A.~Maloney, and S.~Matsuura, ``{Holographic Phases of Renyi
  Entropies},'' {\em JHEP} {\bf 12} (2013) 050,
  \href{http://xxx.lanl.gov/abs/1306.2640}{{\tt 1306.2640}}.

\bibitem{Belin:2014mva}
A.~Belin, L.-Y. Hung, A.~Maloney, and S.~Matsuura, ``{Charged Renyi entropies
  and holographic superconductors},'' {\em JHEP} {\bf 01} (2015) 059,
  \href{http://xxx.lanl.gov/abs/1407.5630}{{\tt 1407.5630}}.

\bibitem{Metlitski:2009iyg}
M.~A. Metlitski, C.~A. Fuertes, and S.~Sachdev, ``{Entanglement Entropy in the
  O(N) model},'' {\em Phys. Rev.} {\bf B80} (2009), no.~11 115122,
  \href{http://xxx.lanl.gov/abs/0904.4477}{{\tt 0904.4477}}.

\bibitem{Ryu:2006bv}
S.~Ryu and T.~Takayanagi, ``{Holographic derivation of entanglement entropy
  from AdS/CFT},'' {\em Phys. Rev. Lett.} {\bf 96} (2006) 181602,
  \href{http://xxx.lanl.gov/abs/hep-th/0603001}{{\tt hep-th/0603001}}.

\bibitem{Faulkner:2013yia}
T.~Faulkner, ``{The Entanglement Renyi Entropies of Disjoint Intervals in
  AdS/CFT},'' \href{http://xxx.lanl.gov/abs/1303.7221}{{\tt 1303.7221}}.

\bibitem{Hartman:2013mia}
T.~Hartman, ``{Entanglement Entropy at Large Central Charge},''
  \href{http://xxx.lanl.gov/abs/1303.6955}{{\tt 1303.6955}}.

\bibitem{Lewkowycz:2013nqa}
A.~Lewkowycz and J.~Maldacena, ``{Generalized gravitational entropy},'' {\em
  JHEP} {\bf 08} (2013) 090, \href{http://xxx.lanl.gov/abs/1304.4926}{{\tt
  1304.4926}}.

\bibitem{Hartman:2014oaa}
T.~Hartman, C.~A. Keller, and B.~Stoica, ``{Universal Spectrum of 2d Conformal
  Field Theory in the Large c Limit},'' {\em JHEP} {\bf 09} (2014) 118,
  \href{http://xxx.lanl.gov/abs/1405.5137}{{\tt 1405.5137}}.

\bibitem{Maxfield:2014kra}
H.~Maxfield, ``{Entanglement entropy in three dimensional gravity},'' {\em
  JHEP} {\bf 04} (2015) 031, \href{http://xxx.lanl.gov/abs/1412.0687}{{\tt
  1412.0687}}.

\bibitem{Skenderis:2009ju}
K.~Skenderis and B.~C. van Rees, ``{Holography and wormholes in 2+1
  dimensions},'' {\em Commun. Math. Phys.} {\bf 301} (2011) 583--626,
  \href{http://xxx.lanl.gov/abs/0912.2090}{{\tt 0912.2090}}.

\bibitem{Maloney:2015ina}
A.~Maloney, ``{Geometric Microstates for the Three Dimensional Black Hole?},''
  \href{http://xxx.lanl.gov/abs/1508.04079}{{\tt 1508.04079}}.

\bibitem{Bizon:2013xha}
P.~Bizoń and J.~Jałmużna, ``{Globally regular instability of $AdS_3$},''
  {\em Phys. Rev. Lett.} {\bf 111} (2013), no.~4 041102,
  \href{http://xxx.lanl.gov/abs/1306.0317}{{\tt 1306.0317}}.

\bibitem{Belin:2017nze}
A.~Belin, C.~A. Keller, and I.~G. Zadeh, ``{Genus Two Partition Functions and
  Renyi Entropies of Large c CFTs},''
  \href{http://xxx.lanl.gov/abs/1704.08250}{{\tt 1704.08250}}.

\bibitem{Cho:2017oxl}
M.~Cho, S.~Collier, and X.~Yin, ``{Recursive Representations of Arbitrary
  Virasoro Conformal Blocks},'' \href{http://xxx.lanl.gov/abs/1703.09805}{{\tt
  1703.09805}}.

\bibitem{Barrella:2013wja}
T.~Barrella, X.~Dong, S.~A. Hartnoll, and V.~L. Martin, ``{Holographic
  entanglement beyond classical gravity},'' {\em JHEP} {\bf 09} (2013) 109,
  \href{http://xxx.lanl.gov/abs/1306.4682}{{\tt 1306.4682}}.

\bibitem{Gaberdiel:2010jf}
M.~R. Gaberdiel, C.~A. Keller, and R.~Volpato, ``{Genus Two Partition Functions
  of Chiral Conformal Field Theories},'' {\em Commun. Num. Theor. Phys.} {\bf
  4} (2010) 295--364, \href{http://xxx.lanl.gov/abs/1002.3371}{{\tt
  1002.3371}}.

\bibitem{zamolodchikov1987conformal}
A.~B. Zamolodchikov, ``Conformal symmetry in two-dimensional space: recursion
  representation of conformal block,'' {\em Theoretical and Mathematical
  Physics} {\bf 73} (1987), no.~1 1088--1093.

\bibitem{Harlow:2011ny}
D.~Harlow, J.~Maltz, and E.~Witten, ``{Analytic Continuation of Liouville
  Theory},'' {\em JHEP} {\bf 12} (2011) 071,
  \href{http://xxx.lanl.gov/abs/1108.4417}{{\tt 1108.4417}}.

\bibitem{zograf1988uniformization}
P.~G. Zograf and L.~A. Takhtadzhyan, ``On uniformization of Riemann surfaces
  and the Weil-Petersson metric on Teichm{\"u}ller and Schottky spaces,'' {\em
  Mathematics of the USSR-Sbornik} {\bf 60} (1988), no.~2 297.

\bibitem{Headrick:2015gba}
M.~Headrick, A.~Maloney, E.~Perlmutter, and I.~G. Zadeh, ``{Rényi entropies,
  the analytic bootstrap, and 3D quantum gravity at higher genus},'' {\em JHEP}
  {\bf 07} (2015) 059, \href{http://xxx.lanl.gov/abs/1503.07111}{{\tt
  1503.07111}}.

\bibitem{Witten:2007kt}
E.~Witten, ``{Three-Dimensional Gravity Revisited},''
  \href{http://xxx.lanl.gov/abs/0706.3359}{{\tt 0706.3359}}.

\bibitem{Calabrese:2009qy}
P.~Calabrese and J.~Cardy, ``{Entanglement entropy and conformal field
  theory},'' {\em J. Phys.} {\bf A42} (2009) 504005,
  \href{http://xxx.lanl.gov/abs/0905.4013}{{\tt 0905.4013}}.

\bibitem{Kraus:2017ezw}
P.~Kraus, A.~Maloney, H.~Maxfield, G.~S. Ng, and J.-q. Wu, ``{Witten Diagrams
  for Torus Conformal Blocks},'' {\em JHEP} {\bf 09} (2017) 149,
  \href{http://xxx.lanl.gov/abs/1706.00047}{{\tt 1706.00047}}.

\bibitem{Zamolodchikov:1985ie}
A.~B. Zamolodchikov, ``{CONFORMAL SYMMETRY IN TWO-DIMENSIONS: AN EXPLICIT
  RECURRENCE FORMULA FOR THE CONFORMAL PARTIAL WAVE AMPLITUDE},'' {\em Commun.
  Math. Phys.} {\bf 96} (1984) 419--422.

\bibitem{Maxfield:2017rkn}
H.~Maxfield, ``{A view of the bulk from the worldline},''
  \href{http://xxx.lanl.gov/abs/1712.00885}{{\tt 1712.00885}}.

\bibitem{Chang:2016ftb}
C.-M. Chang and Y.-H. Lin, ``{Bootstrap, universality and horizons},'' {\em
  JHEP} {\bf 10} (2016) 068, \href{http://xxx.lanl.gov/abs/1604.01774}{{\tt
  1604.01774}}.

\bibitem{Dorn:1994xn}
H.~Dorn and H.~J. Otto, ``{Two and three point functions in Liouville
  theory},'' {\em Nucl. Phys.} {\bf B429} (1994) 375--388,
  \href{http://xxx.lanl.gov/abs/hep-th/9403141}{{\tt hep-th/9403141}}.

\bibitem{Zamolodchikov:1995aa}
A.~B. Zamolodchikov and A.~B. Zamolodchikov, ``{Structure constants and
  conformal bootstrap in Liouville field theory},'' {\em Nucl. Phys.} {\bf
  B477} (1996) 577--605, \href{http://xxx.lanl.gov/abs/hep-th/9506136}{{\tt
  hep-th/9506136}}.

\bibitem{Klebanov:1999tb}
I.~R. Klebanov and E.~Witten, ``{AdS / CFT correspondence and symmetry
  breaking},'' {\em Nucl. Phys.} {\bf B556} (1999) 89--114,
  \href{http://xxx.lanl.gov/abs/hep-th/9905104}{{\tt hep-th/9905104}}.

\bibitem{sullivan1979density}
D.~Sullivan, ``The density at infinity of a discrete group of hyperbolic
  motions,'' {\em Inst. Hautes {\'E}tudes Sci. Publ. Math} {\bf 50} (1979),
  no.~2979 171--202.

\bibitem{patterson1989selberg}
S.~Patterson, ``The Selberg zeta-function of a Kleinian group,'' in {\em Number
  theory, trace formulas and discrete groups}, pp.~409--441.
\newblock Elsevier, 1989.

\bibitem{mumford2002indra}
D.~Mumford, C.~Series, and D.~Wright, {\em Indra's pearls: the vision of Felix
  Klein}.
\newblock Cambridge University Press, 2002.

\bibitem{patterson1976limit}
S.~J. Patterson, ``The limit set of a Fuchsian group,'' {\em Acta mathematica}
  {\bf 136} (1976), no.~1 241--273.

\bibitem{sullivan1987related}
D.~Sullivan {\em et.~al.}, ``Related aspects of positivity in Riemannian
  geometry,'' {\em Journal of differential geometry} {\bf 25} (1987), no.~3
  327--351.

\bibitem{bishop1997hausdorff}
C.~J. Bishop and P.~W. Jones, ``Hausdorff dimension and Kleinian groups,'' {\em
  Acta Mathematica} {\bf 179} (1997), no.~1 1--39.

\bibitem{Giombi:2008vd}
S.~Giombi, A.~Maloney, and X.~Yin, ``{One-loop Partition Functions of 3D
  Gravity},'' {\em JHEP} {\bf 08} (2008) 007,
  \href{http://xxx.lanl.gov/abs/0804.1773}{{\tt 0804.1773}}.

\bibitem{patterson2001divisor}
S.~J. Patterson, P.~A. Perry, {\em et.~al.}, ``The divisor of Selberg's zeta
  function for Kleinian groups,'' {\em Duke Mathematical Journal} {\bf 106}
  (2001), no.~2 321--390.

\bibitem{mcmullen3}
C.~T. McMullen, ``Hausdorff dimension and conformal dynamics, III: Computation
  of dimension,'' {\em American journal of mathematics} (1998) 691--721.

\bibitem{mcmullen1}
C.~T. McMullen, ``Hausdorff dimension and conformal dynamics I: Strong
  convergence of Kleinian groups,''.

\bibitem{dodziuk1987estimating}
J.~Dodziuk, T.~Pignataro, B.~Randol, and D.~Sullivan, ``Estimating small
  eigenvalues of Riemann surfaces,'' {\em The legacy of Sonya Kovalevskaya
  (Cambridge, Mass., and Amherst, Mass., 1985), Contemp. Math} {\bf 64} (1987)
  93--121.

\bibitem{Calabrese:2012ew}
P.~Calabrese, J.~Cardy, and E.~Tonni, ``{Entanglement negativity in quantum
  field theory},'' {\em Phys. Rev. Lett.} {\bf 109} (2012) 130502,
  \href{http://xxx.lanl.gov/abs/1206.3092}{{\tt 1206.3092}}.

\bibitem{hou2016smooth}
Y.~Hou, ``On smooth moduli space of Riemann surfaces,'' {\em arXiv preprint
  arXiv:1610.03132} (2016).

\end{thebibliography}\endgroup

\end{document}